\documentclass[reprint,pra,aps,amsmath,amssymb,floatfix]{revtex4-1}

\usepackage{dsfont}
\usepackage[caption=false]{subfig}

\newcommand{\subsubfloat}[2]{%
  \begin{tabular}{@{}c@{}}#1\\#2\end{tabular}%
}

\usepackage{float}
\usepackage{graphicx}
\usepackage{dcolumn}
\usepackage{bm}
\usepackage{color}
\usepackage{lipsum}

\usepackage{hyperref}

\usepackage{bbm}

\DeclareMathOperator{\Tr}{Tr}

\newcommand{\eq}[1]{(\ref{eq:#1})}  
\newcommand{\fig}[1]{{\ref{fig:#1}}}

\newcommand{\Sec}[1]{\textbf{\ref{sec:#1}}} 

\newcommand{\cd}{c^\dagger}
\newcommand{\pd}{{\phantom{\dagger}}}
\newcommand{\cc}{c^\pd}
\newcommand{\ii}{\textrm{i}}
\newcommand{\dd}{\textrm{d}}

\begin{document}

\title{Complexity and Floquet dynamics: non-equilibrium Ising phase transitions}

\author{Giancarlo Camilo}
\email{gcamilo@iip.ufrn.br}
\affiliation{International Institute of Physics, Universidade Federal do Rio Grande do Norte\\
Campus Universit\'ario, Lagoa Nova, Natal-RN 59078-970, Brazil}

\author{Daniel Teixeira}
\email{dteixeira@usp.br}
\affiliation{Institute of Physics, University of S\~ao Paulo, 05314-970 S\~ao Paulo, Brazil.}

\date{\today}

\begin{abstract}

We study the time-dependent circuit complexity of the periodically driven transverse field Ising model using Nielsen's geometric approach. In the high-frequency driving limit the system is known to exhibit non-equilibrium phase transitions governed by the amplitude of the driving field. We analytically compute the complexity in this regime and show that it clearly distinguishes between the different phases, exhibiting a universal linear behavior at early times. We also evaluate the time averaged complexity, provide evidence of non-analytic behavior at the critical points, and discuss its origin. 
Finally, we comment on the freezing of quantum dynamics at specific configurations and on the use of complexity as a new tool to understand quantum phase transitions in Floquet systems.

\end{abstract}

\maketitle

\section{\label{sec:intro} Introduction\protect }

Understanding the organizing principles underlying the non-equilibrium dynamics of quantum many-body systems is of key importance for the development of new quantum materials. The concept of universality, which provides a unified description of equilibrium critical phenomena, is not well understood for systems far from equilibrium. 
In the case of adiabatic dynamics the so-called Kibble-Zurek mechanism and its quantum extension can provide some insights into the breakdown of adiabaticity close to a quantum phase transition (QPT) point and the associated scaling behavior of the excitation density of defects \cite{Kibble:1976sj,Zurek:1985qw,PhysRevLett.95.245701,PhysRevB.72.161201,PhysRevLett.95.105701},
which opened a venue for the analysis of universal features in non-equilibrium QPT. The preclusion of adiabatically connecting states belonging to different quantum phases can be given a geometric interpretation as a diverging curvature with the introduction of a metric on the Hilbert space \cite{GU_2010}. 
This geometric paradigm is part of an ongoing effort in the last two decades to employ concepts and tools from quantum information science to improve our understanding of quantum many-body physics. 
This approach has led to remarkable progresses, such as the discovery of topologically ordered states
and of the critical behavior of entanglement close to a QPT \cite{Osterloh_2002,Vidal_2003} 
(see \cite{zeng2015quantum} for a review). 

As part of this effort, \cite{Liu_2020,PhysRevB.101.174305,jaiswal2020complexity} proposed to characterize QPTs, including topological ones, using a geometric notion of circuit complexity introduced by Nielsen \cite{nielsen2005geometric,Nielsen_2006}. Inspired by its computer science analogue, this object quantifies how difficult it is to construct a particular unitary operator that maps between a pair of given reference and target states, i.e., the minimum number of basic operations needed to implement this task. With an appropriate definition of depth functional associated with each circuit, the space of allowed unitaries acquires a Riemannian structure and the problem of finding the optimal circuit reduces to finding minimal geodesics in this geometry. Nielsen's complexity has recently also attracted a lot of interest from the high energy physics community due to conjectured connections with black hole properties within the scope of the holographic duality \cite{Stanford:2014jda,Brown:2015bva,Jefferson:2017sdb,Caputa:2018kdj,Balasubramanian:2019wgd}.

A major difficulty to unravel universal non-equilibrium properties independent of specific models
comes from the variety of ways in which a system can be put away from equilibrium. 
Perhaps the simplest and one of the most studied among these non-equilibrium protocols is that of a quantum quench, where a parameter of the Hamiltonian is suddenly changed and the system is let evolve under the new Hamiltonian \cite{Polkovnikov_2011,Mitra_2018}. 
This also includes the study of the quench dynamics of circuit complexity \cite{Alves:2018qfv,Camargo:2018eof,Liu:2019qyx,Liu_2020,PhysRevB.101.174305,ali2018postquench}. 
Here we propose to go a step further in the endeavour of using circuit complexity as a novel tool to understand the dynamics of quantum many-body systems and explore a different non-equilibrium protocol corresponding to the periodic driving of many-body systems. These so-called Floquet systems can be experimentally realized with ultracold quantum gases in optical lattices (see \cite{Eckardt_2017,Oka_2019} for a review of theoretical and experimental results) and give rise to several exotic phenomena such as dynamical localization, Floquet topological insulators, and driving-induced phase transitions \cite{Eckardt_2017,Oka_2019,PhysRevLett.99.220403,Lindner_2011,Jotzu_2014}.

We shall focus on the Ising model under periodic driving of the transverse field 
\cite{Mukherjee_2009,Das_2010,Russomanno_2012,Bastidas_2012}. The model can be solved analytically in the fast driving limit, where it is effectively described by a time-independent Hamiltonian and it displays quantum phase transitions of non-equilibrium nature controlled by the transverse field amplitudes. It also exhibits the phenomenon of dynamic localization \cite{PhysRevB.34.3625}, where the time evolution gets frozen to the initial state as a consequence of a many-body version of the coherent destruction of tunneling (CDT) \cite{PhysRevLett.67.516} that occurs in each momentum sector of the Hilbert space.
CDT has been observed experimentally 
and it is particularly important for quantum dynamics control 
\cite{2008PhRvL.100s0405K,2005PhRvL..95z0404E,2006PhRvL..96r7002S}.

In this setup, we compute the circuit complexity of the instantaneous time-evolved state and argue that it can be used to characterize these non-equilibrium phase transitions, showing that its time-average exhibits non-analytic behavior at the critical points. We also unveil a universal linear behavior at early times and show that the CDT phenomenon is naturally diagnosed by points of vanishing complexity. Our work takes to a next level the connection between circuit complexity and quantum phase transitions, opening the route for periodically driven systems and dynamical phase transitions.

\section{\label{sec:level1} The driven transverse field Ising model \protect }

We consider a periodically driven transverse field Ising model (TFIM) described by the Hamiltonian
\begin{equation}\label{eq:H}
H(t) = -J\sum_{i=1}^L\sigma_i^z\sigma_{i+1}^z -g(t)\sum_{i=1}^L\sigma_i^x ,
\end{equation}
where $\sigma_i^\alpha$ are Pauli matrices at the $i$-th lattice site, $J>0$ is the exchange coupling, and $g(t) = g_0 + g_1\cos{\Omega t}$ is the transverse field, made of a constant contribution $g_0$ and a monochromatic driving with frequency $\Omega$. Here we assume a closed lattice with periodic boundary conditions $\sigma_{L+1}^\alpha\equiv\sigma_{1}^\alpha$ and restrict to even $L$. The $\mathbb{Z}_2$ symmetry of the model is implemented by the parity operator $\mathcal{P}=\prod_{i=1}^L\sigma_i^x$, 
resulting in a decomposition of the Hilbert space into a direct sum of parity odd ($\mathcal{P}=-1$) or even ($\mathcal{P}=+1$) subspaces \cite{LIEB1961407}, each of dimension $2^{L-1}$ -- the so-called Ramond (R) and Neveu-Schwarz (NS) sectors, respectively. We shall focus on the NS sector only. 

In terms of Jordan-Wigner fermions $c_j$, after the discrete Fourier transform, $\cc_j=\frac{e^{-\ii\,\pi/4}}{\sqrt{L}}\sum_{k\in\text{BZ}}\cc_ke^{\ii kj}$, the Hamiltonian can be written as $H(t) = \sum_{k>0} H_k(t)$ with
\begin{eqnarray}\label{eq:Hk}
 H_k(t) &=& [2g(t)-\omega_k](\cd_k\cc_{k}+\cd_{-k}\cc_{-k}) + \notag \\
 &+&\Delta_k(\cd_k\cd_{-k}+\cc_{-k}\cc_k)-\omega_k,
\end{eqnarray}
where $\omega_k=2J\cos k,\Delta_k=2J\sin k$, and we have neglected the trivial contribution $-2Lg(t)$. The momenta are constrained to the first Brillouin zone, $\text{BZ}=\{\pm\frac{\pi}{L},\pm\frac{3\pi}{L},\ldots,\pm\frac{(L-1)\pi}{L}\}$ by the antiperiodic boundary condition satisfied by the $c_j$ in the NS sector. 

Since the Hamiltonian conserves momentum and parity, the state of a system initialized in a ground state of the undriven model will acquire at any time $t$ the following form 
\cite{Dziarmaga_2005,Mukherjee_2009,Bhattacharyya_2012},
\begin{equation}\label{eq:BCS}
    \vert \psi(t)\rangle = \bigotimes_{k>0}\big[u_k(t)\,\vert 1_{-k}1_k \rangle + v_k(t)\,\vert 0_{-k}0_k \rangle \big]\,,
\end{equation}
that is, for each $k$ the dynamics is restricted to the two-level Nambu subspace spanned by $\{\vert 0_{-k}0_k \rangle,\vert 1_{-k}1_k \rangle\}$. One can unify the coefficients into the spinor $\Psi_k(t)\equiv(u_k(t) ~ v_k(t))^{\intercal}$, which obeys Schrödinger equation generated by the Bogoliubov–de Gennes (BdG) Hamiltonian \eq{Hk}, such that the dynamics of each momentum mode takes the form of a driven two-level system. In terms of $\mathfrak{u}(2)$ generators, one has
\begin{equation}\label{eq:HBdG}
    H_k(t) = \big(2g(t)-\omega_k\big)\,\sigma_k^z +\Delta_k\,\sigma_k^x-\omega_k\,\mathds{1}_k \, .
\end{equation}

According to the Floquet theorem, the solution can be written as
\begin{equation}\label{eq:Psi_sol}
\Psi_k(t) = \sum_{\lambda=\pm}A_\lambda e^{\ii\,\varepsilon_k^{(\lambda)}t}\Phi^{(\lambda)}_k(t)\,,
\end{equation}
where the Floquet modes $\Phi^{(\pm)}_k(t)=\Phi^{(\pm)}_k(t+ 2\pi\Omega^{-1})$ are periodic with the same period of the external driving and satisfy the time-independent Schrödinger equation
for the Floquet Hamiltonian $\mathcal{H}_k\equiv H_k(t)-\ii\,\partial_t$. The Floquet quasienergies $\varepsilon_k^{(\lambda)}$ are only defined modulo $\Omega$ since $e^{\ii\,m\Omega t}\,\Phi^{(\pm)}_k(t)$ for any $m\in\mathbb{Z}$ obviously defines another Floquet mode with quasienergy shifted by $m\Omega$, meaning in particular that the driven system admits no notion of ground state. 
In the following section we discuss exact solutions in the limit of high driving frequency following the approach of \cite{Ashhab_2007,Bastidas_2012}.

\subsection{\label{sec:level2} High-frequency driving approximation}

It will be convenient to split the constant part of the transverse field as $g_0=\delta g_0^\pd+\tilde{g_0}$, where $\tilde{g}_0$ is a resonant value (to be determined) and $\delta g_0^\pd\equiv g_0-\tilde{g}_0$ is a detuning measuring the distance to this resonance. 

The dynamics can then be solved by going to a rotating frame tweaked to the driving field $g(t)$ through a unitary transformation ${\cal R}_k(t)$. First we split $H_k(t)=H_{k}^0(t)+H_{k}
^1$ with $H_{k}^0(t)\equiv2(\tilde{g}_0+g_1\cos\Omega t)\,\sigma_k^z$ and $H_{k}^1\equiv\Delta_k\,\sigma_k^x+(2\delta g_0-\omega_k)\,\sigma_k^z-\omega_k\,\mathds{1}_k$  and go to the interaction picture with $H_{k}^1$ as the interaction Hamiltonian. The desired transformation is the time evolution operator associated with $H_{k}^0(t)$, namely 
\begin{equation}
{\cal R}_k(t) = e^{-\frac{\ii}{2}\alpha(t)\sigma^z_k}, \quad \alpha(t) = 4\tilde{g}_0t + \frac{4g_1}{\Omega}\sin\Omega t\, . 
\end{equation}
The rotated Hamiltonian $\tilde{H}_{k}(t)\equiv {\cal R}_k^\dagger(t)H^\pd_{k}(t){\cal R}^\pd_k(t)=H_{k}^0(t)+\tilde{H}_{k}^{1}(t)$ has the same free contribution and a rotating part given by 
\begin{equation}\label{eq:H1_int}
\tilde{H}_{k}^{1}(t) = 
\begin{pmatrix}
2\delta g_0-2\omega_k & \Delta_k e^{\ii\,\alpha(t)}\\
\Delta_k e^{-\ii\,\alpha(t)} & -2\delta g_0
\end{pmatrix}.
\end{equation}
States $\vert\psi(t)\rangle_k$ whose dynamics is governed by \eq{HBdG} are mapped to rotated states $\vert\tilde{\psi}(t)\rangle^\pd_k={\cal R}^{\dag}_k(t)\,\vert\psi(t)\rangle^\pd_k$ with Schrödinger time evolution dictated by \eq{H1_int}. The full rotated Hamiltonian $\tilde{H}^{1}(t)\equiv\sum_{k>0}\tilde{H}_{k}^{1}(t)$ in terms of the original spins contains all possible nearest-neighbor free fermion terms $\sigma_i^x, \sigma_{i}^z\sigma_{i+1}^z, \sigma_{i}^y\sigma_{i+1}^y, \sigma_{i}^z\sigma_{i+1}^y, \sigma_{i}^y\sigma_{i+1}^z$ \cite{Bastidas_2012}.

In order to determine the resonance condition, we first make use of the Jacobi-Anger expansion 
$ e^{\ii\, z\sin\Omega t} = \sum_{n\in\mathbb{Z}} \mathcal{J}_n(z)\exp(\ii \,n\Omega t),$
where $\mathcal{J}_n(z)$ are Bessel functions of the first kind, to rewrite (\ref{eq:H1_int}) in the form $\tilde{H}^{1}_k(t)=\sum_{n\in\mathbb{Z}}\tilde{h}^{(n)}e^{\ii(4\tilde{g}_0-n\Omega)t}$ for some $\tilde{h}^{(n)}$. 
Then, the high-frequency approximation (sometimes referred as rotating wave approximation) is performed assuming that all the terms in the summation oscillate wildly and can be neglected with respect to a single resonant term given by
\begin{equation}\label{eq:g0resonance}
\tilde{g}_0^{(\ell)}=\ell\frac{\Omega}{4}, \quad \ell\in\mathbb{Z}\,.
\end{equation}
The corresponding detuning parameter will be denoted by $\delta g_0^{(\ell)}=g_0^\pd-\tilde{g}_0^{(\ell)}$. 
As a result, the effective Hamiltonian describing the dynamics of the system at the $\ell$-th resonance becomes time-independent.

In terms of the original spins, the full rotating frame Hamiltonian $\tilde{H}^{1(\ell)} = \sum_{k>0}\tilde{H}_{k}^{{1}(\ell)}$ takes the form
\begin{equation}\label{eq:H_XY}
\tilde{H}^{1(\ell)} = -\sum_{j=1}^L\left[\delta g_0^{(\ell)}\sigma_j^x + J_+^{(\ell)}\sigma_j^z \sigma_{j+1}^z + J_-^{(\ell)}\sigma_j^y \sigma_{j+1}^y\right]
\end{equation}
with $J_\pm^{(\ell)}\equiv\frac{J}{2}(1\pm\gamma^{(\ell)})$ and
\begin{equation}\label{eq:anisotropy}
\gamma^{(\ell)} \equiv (-1)^\ell\mathcal{J}_{\ell}\left(\frac{4g_1}{\Omega}\right).
\end{equation}
This is unitarily equivalent to the familiar transverse XY chain with anisotropy parameter $\gamma^{(\ell)}$ \cite{LIEB1961407,PhysRevA.2.1075}. The non-trivial dependence of $\gamma^{(\ell)}$ on $\ell,\Omega,g_1$ already anticipates the influence of the driving on the critical behavior of the system, to be confirmed in the next section. 
Near the resonance there is pure coupling between the two-level system basis states at $\min_k \vert\omega_k\vert$, with oscillation frequency given by $\omega_{\rm eff} 
= J\vert\gamma^{(\ell)}\vert$, indicating that this large-$\Omega$ approximation remains valid as long as $\delta g_0^{(\ell)}, \omega_{\rm eff} \ll \Omega$.

\subsection{Nonequilibrium QPTs in the rotating frame}

The XY model \eq{H_XY} describing the high-$\Omega$ dynamics in the rotating frame is exactly solvable via Jordan-Wigner and discrete Fourier transforms following closely the discussion for the TFIM in Section \Sec{level1}. The Bogoliubov angle $\vartheta_{k,\ell}$ defined by
\begin{equation}\label{eq:thetak}
\tan(2\vartheta_{k,\ell})=\frac{\Delta_k\,\gamma^{(\ell)}}{2\delta g_0^{(\ell)}-\omega_k}
\end{equation}
diagonalizes the corresponding BdG Hamiltonian to the free fermion form $\tilde{H}^{1(\ell)} = \sum_{k>0} \epsilon_{k,\ell}\left(b^\dagger_k b^\pd_k-\frac{1}{2}\right)$ 
with
\begin{equation}
    \epsilon_{k,\ell} = \sqrt{\big(2\delta g_0^{(\ell)}-\omega_{k}\big)^2+\big(\Delta_{k}\,\gamma^{(\ell)}\big)^2}\,.
    \label{eq:Floquet-spectrum}
\end{equation}
The positive and negative energy eigenstates, with eigenvalues $\epsilon_{k,\ell}^\pm=-\omega_k\pm\epsilon_{k,\ell}$, are $\phi_{k,+}^{(\ell)} = (\cos\vartheta_{k,\ell}~-\sin\vartheta_{k,\ell})^{\intercal}$ and $\phi_{k,-}^{(\ell)} = (\sin\vartheta_{k,\ell}~\cos\vartheta_{k,\ell})^{\intercal}$.

\begin{figure}
    \centering
    \includegraphics[scale=0.27]{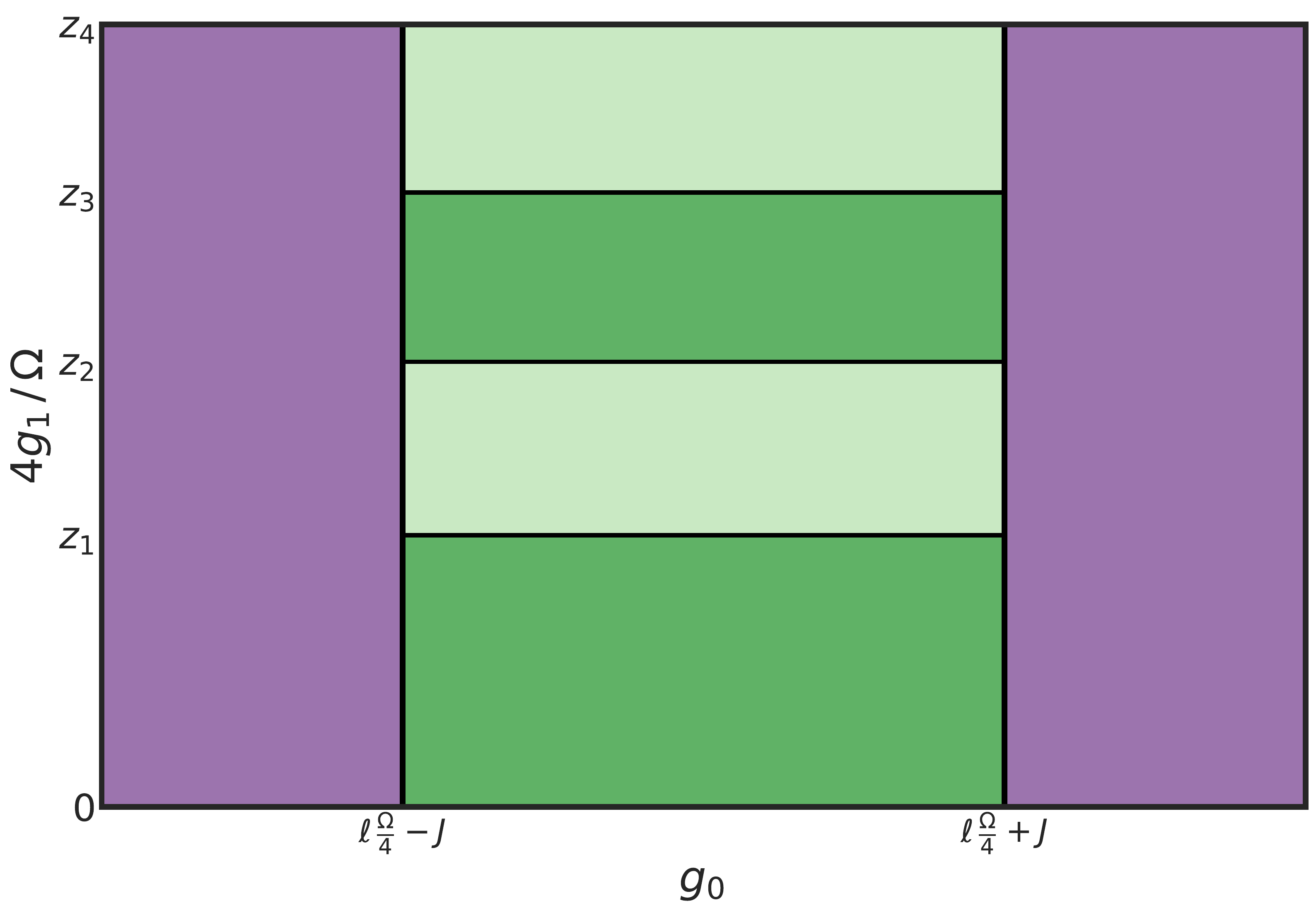}
    \caption{Non-equilibrium phase diagram as a function of the transverse field strengths $g_0,g_1$ in the high-$\Omega$ regime. The phases are PM (purple), FMZ (green) and FMY (light green). Vertical and horizontal lines identify the Ising-like and anisotropic phase transitions, respectively. The latter are located at $z_i$, the $i$-th root of $\mathcal{J}_\ell(z)$ ($\ell=2$ shown in the plot); the width $z_{i+1}-z_{i}$ quickly approaches $\pi$ as $i$ grows. }
    \label{fig:phasediagram}
\end{figure}

The model is known to present two critical lines: an Ising-like QPT between a ferromagnetic and a paramagnetic phase at $\vert\delta g_0^{(\ell)}\vert = J$; and an anisotropic QPT at $\gamma^{(\ell)} = 0$ (provided that $\vert \delta g_0^{(\ell)}\vert <J$) between two distinct phases FMY ($\gamma^{(\ell)}<0$) and FMZ ($\gamma^{(\ell)}>0$) with ferromagnetic order along the $y$ and $z$ directions, respectively. For a given $\ell$, the former defines a pair of lines $\delta g_0^{(\ell)} = \pm J$ while the latter corresponds to an infinite family of critical lines, one for each zero of $\mathcal{J}_\ell(z)$. The phase diagram as a function of the transverse field strengths $g_0,g_1$ for fixed (and large) $\Omega$ is illustrated in Figure \fig{phasediagram}. 
The FMY-FMZ transition lines are almost evenly spaced (except for the first few) since the sequence $\{z_{i+1}-z_{i}\}_{i\in\mathbb{Z}^+}$ of differences between two subsequent Bessel zeros converges very quickly to the constant value $\pi$, as seen intuitively from the asymptotic behavior  $\mathcal{J}_\ell(z)\approx\sqrt{\frac{2}{\pi z}}\cos\left[z-(2\ell+1)\frac{\pi}{4}\right]$ at $z\gg \ell$. 
Note that in the special case $\delta g_0^{(\ell)} = 0$, i.e., when $g_0$ is tuned exactly to the resonant value $g_0^{(\ell)}$, the transverse field in \eq{H_XY} disappears and we are left only with the anisotropic transitions.

The horizontal lines in Figure \fig{phasediagram} occur at $\gamma^{(\ell)}=0$, where $\omega_{\rm eff} = 0$ forces the quantum tunneling between $\sigma^z_k$ eigenstates to completely freeze. This phenomenon, known as coherent destruction of tunneling (CDT), occurs at every sector $k$ once the driving amplitude is fine-tuned to one of the Bessel zeros, leading to a coherent suppression of the dynamics even at infinite $L$. We will show how this dynamic localization effect manifests in the circuit complexity in the next section. 

\subsection{Floquet modes and quasienergies}

The Floquet modes that define a basis for the dynamics in the Schrödinger picture follow by applying ${\cal R}_k$ to the eigenstates of the XY Hamiltonian,
\begin{equation}
\Phi_{k,\pm}^{(\ell)}(t)\equiv e^{-\ii\left(\frac{\ell\Omega}{2} t+\frac{2g_1}{\Omega}\sin{\Omega t}\right)}{\cal R}_k^{(\ell)}(t)\,\phi_{k,\pm}^{(\ell)}    
\end{equation}
(the $U(1)$ phase is added for convenience) and correspond to quasienergies
\begin{equation}
\varepsilon_{k,\ell}^{\pm}\equiv -\omega_k\pm\,\epsilon_{k,\ell} + \frac{\ell\Omega}{2}\,.    
\end{equation}
Here we recall that there is an infinite family of Floquet modes, labelled by an integer $m$ that is omitted here, corresponding to the rescaling $\Phi_{k,\pm}^{(\ell)}(t)\to e^{\ii m\Omega t}\,\Phi_{k,\pm}^{(\ell)}(t)$ and shift $\varepsilon_{k,\ell}^{\pm}\to \varepsilon_{k,\ell}^{\pm}+m\Omega$. We choose the $m=0$ representative without loss of generality.

Finally, the general solution \eq{Psi_sol} with initial condition $\Psi_{k}^{(\ell)}(0)\equiv\big(u_{k}^{(\ell)}(0)~v_{k}^{(\ell)}(0)\big)^{\intercal}$ is completely determined due to the orthogonality of the Floquet modes by the coefficients 
$A_{k,\ell}^\pm\equiv \Psi_{k}^{(\ell)}(0)\,\phi_{k,\pm}^{(\ell)}$. 
We will focus on $\Psi_{k}^{(\ell)}(0)=\big(0~1\big)^{\intercal}$, i.e., a system initialized in the paramagnetic state of the undriven model with all the spins aligned along the $x$ direction, $\bigotimes_{k>0}\vert 0_{k}0_{-k}\rangle$, corresponding to $A_{k,\ell}^+=-\sin\vartheta_{k,\ell}$ and $A_{k,\ell}^-=\cos\vartheta_{k,\ell}$. In terms of the spinor components introduced in \eq{BCS}, the explicit solution $\Psi_k^{(\ell)}(t)$ reads (up to a global phase $e^{-\ii\,\varepsilon_{k,\ell}^{-}t}$)
\begin{equation}\label{eq:uvsol}
\begin{pmatrix}
u_{k}^{(\ell)}(t)\\
v_{k}^{(\ell)}(t)
\end{pmatrix}
=
\begin{pmatrix}
e^{-\ii\,\alpha^{(\ell)}(t)}\left(1-e^{-2\ii\,\epsilon_{k,\ell}t}\right)\sin\vartheta_{k,\ell}\cos\vartheta_{k,\ell}\\
\cos^2\vartheta_{k,\ell}+e^{-2\ii\,\epsilon_{k,\ell}t}\,\sin^2\vartheta_{k,\ell}
\end{pmatrix}
.
\end{equation}

\section{Complexity across nonequilibrium QPTs}

In this Section we discuss the circuit complexity of the instantaneous states \eq{BCS} using the geometric approach introduced in \cite{nielsen2005geometric,Nielsen_2006}. Namely, we look for the optimal circuit $U=U(t)$ connecting the reference and target states, $\vert T\rangle=U\vert R\rangle$, with $\vert R\rangle=\bigotimes_{k>0}\vert 0_{k}0_{-k}\rangle$ and $\vert T\rangle=\vert \Psi(t)\rangle$. Note that we choose $\vert R\rangle$ to be the same as the initial condition $\vert \Psi(0)\rangle$ so that the complexity starts from a vanishing value at $t=0$.
Factorization of states in fixed-momentum sectors implies that $U=\bigotimes_{k>0}U_k$. In terms of Nambu spinors, each admissible $U_k$ is a Bogoliubov transformation taking the reference spinor $\Psi_k^R=(0 ~ 1)^{\intercal}$ to $\Psi_k^{(\ell)}(t)=(u_k^{(\ell)}(t) ~ v_k^{(\ell)}(t))^{\intercal}$ derived in \eq{uvsol}. Since these are $SU(2)$ transformations, it is natural to seek for factorized circuits $\mathcal{U}(s)=\bigotimes_{k>0}\mathcal{U}_k(s)$ with each factor having the Hamiltonian form
\begin{equation}\label{eq:Ukpathordered}
\mathcal{U}_k(s) = \mathcal{P}e^{\int_0^sH_k(s')\dd s'},\quad H_k(s')\equiv\sum_IY_k^I(s'){\cal O}_I,
\end{equation}
where $s\in[0,1]$ is a continuous parameter, the functions $Y_k^I(s)=-\frac{1}{2}\Tr\left[\partial_s\,\mathcal{U}_k(s)\,\mathcal{U}_k(s)^{-1}{\cal O}_I\right]$ identify a particular circuit, ${\cal O}_I\in\{\ii\,\sigma^x,\ii\,\sigma^y,\ii\,\sigma^z\}$ are the $\mathfrak{su}(2)$ generators (our fundamental gates), and $\mathcal{P}$ a path-ordering operator ensuring that the circuit is built from smaller to larger values of $s$. The boundary conditions $\mathcal{U}_k(s=0)=\mathds{1}$ and $\mathcal{U}_k(s=1)=U_k$ guarantee that any such circuit implements the desired task of connecting the two given states. The optimal circuit is found by minimizing an associated depth functional, $\mathcal{D}[\mathcal{U}_k]=\int_0^s \dd s'F\big(\{Y_k(s')\}\big)$, and the corresponding complexity corresponds to the depth of this optimal circuit,
\begin{equation}
\mathcal{C}[\mathcal{U}_k] = \min_{\{Y_k^I(s)\}}\mathcal{D}[\mathcal{U}_k] = \mathcal{D}[\mathcal{U}_k^{\text{opt}}]\,.
\end{equation}
We choose as cost function $F$ the Euclidean norm $F\big(\{Y_k\}\big)=\left(\sum_I|Y_k^I|^2\right)^{1/2}$, which is the simplest one satisfying all the required properties from complexity measures \cite{nielsen2005geometric} (see \cite{Jefferson:2017sdb} for alternatives).

To solve the minimization problem it will be convenient to use the polar representation of the components of \eq{uvsol},
namely $v_{k}^{(\ell)}(t) \equiv \cos\Theta_{k,\ell}(t) \, e^{\ii\,\varphi^v_{k,\ell}(t)}$ and  $u_{k}^{(\ell)}(t) \equiv \sin\Theta_{k,\ell}(t) \, e^{\ii\,\varphi^u_{k,\ell}(t)}$ with $0 \leq \Theta_{k,\ell}(t) \leq \pi/2$, and further discarding a global phase to choose the element in the ray of $\Psi_k^{(\ell)}(t)$ to be
\begin{equation}\label{eq:targetspinor}
\Psi_{k}^{(\ell)}(t) = 
\begin{pmatrix}
e^{\ii\,\beta_{k,\ell}(t)}\sin\Theta_{k,\ell}(t)\\
\cos\Theta_{k,\ell}(t)
\end{pmatrix}
\end{equation}
with $\beta_{k,\ell}(t)\equiv \varphi^u_{k,\ell}(t)-\varphi^v_{k,\ell}(t)$.
The Bogoliubov transformation to be implemented thus assumes the form
\begin{equation}
U_k=\begin{pmatrix}
\cos\Theta_{k,\ell}(t) & e^{\ii\,\beta_{k,\ell}(t)}\sin\Theta_{k,\ell}(t) \\
-e^{-\ii\,\beta_{k,\ell}(t)}\sin\Theta_{k,\ell}(t) & \cos\Theta_{k,\ell}(t)
\end{pmatrix}
\,.
\end{equation}

This suggests a parametrization of the circuit $\mathcal{U}_k(s) \in SU(2)$ for each momentum sector $k$ in terms of Hopf coordinates ($\phi_1,\phi_2,\omega$),
\begin{equation}
\mathcal{U}_k(s) = 
\begin{pmatrix}
e^{\ii\,\phi_1(s)}\cos\omega(s) & e^{\ii\,\phi_2(s)}\sin\omega(s)\\
-e^{-\ii\,\phi_2(s)}\sin\omega(s)& e^{-\ii\,\phi_1(s)}\cos\omega(s)
\end{pmatrix}
\,.
\end{equation}
With this at hand, it is straightforward to show that the optimal circuit minimizes the functional 
\begin{equation}
\mathcal{D}[\mathcal{U}_k] = \int_0^1\dd s' \sqrt{\omega'^2+\cos^2\omega\,\phi_1'^2+\sin^2\omega\,\phi_2'^2}\,.
\end{equation}
The minimum corresponds to constant phase functions $\phi_1(s)=\phi_1^0,\phi_2(s)=\phi_2^0$ and the linear profile $\omega(s)=\omega_0+s\,\omega_1$, which immediately implies $\mathcal{D}[\mathcal{U}_k^{\text{opt}}]=|\omega_1|$. The boundary condition at $s=0$ then fixes $\phi_1^0=0$ and $\omega_0=0$, while the one at $s=1$ fixes $\omega_1=\Theta_{k,\ell}(t)$ and $\phi_2^0=\beta_{k,\ell}(t)$. Putting all together and summing over all momentum sectors we obtain the circuit complexity $\mathcal{C}(t) = \sum_{k>0} \left|\Theta_{k,\ell}(t)\right|$ or, explicitly,
\begin{equation}\label{eq:cpxt}
\mathcal{C}(t) = 
\sum_{k>0} \left\vert\arcsin\left(\frac{\Delta_k\gamma^{(\ell)}}{\epsilon_{k,\ell}}\sin(\epsilon_{k,\ell}t)\right)\right\vert.
\end{equation}

\begin{figure}
    \centering
    \includegraphics[scale=0.25]{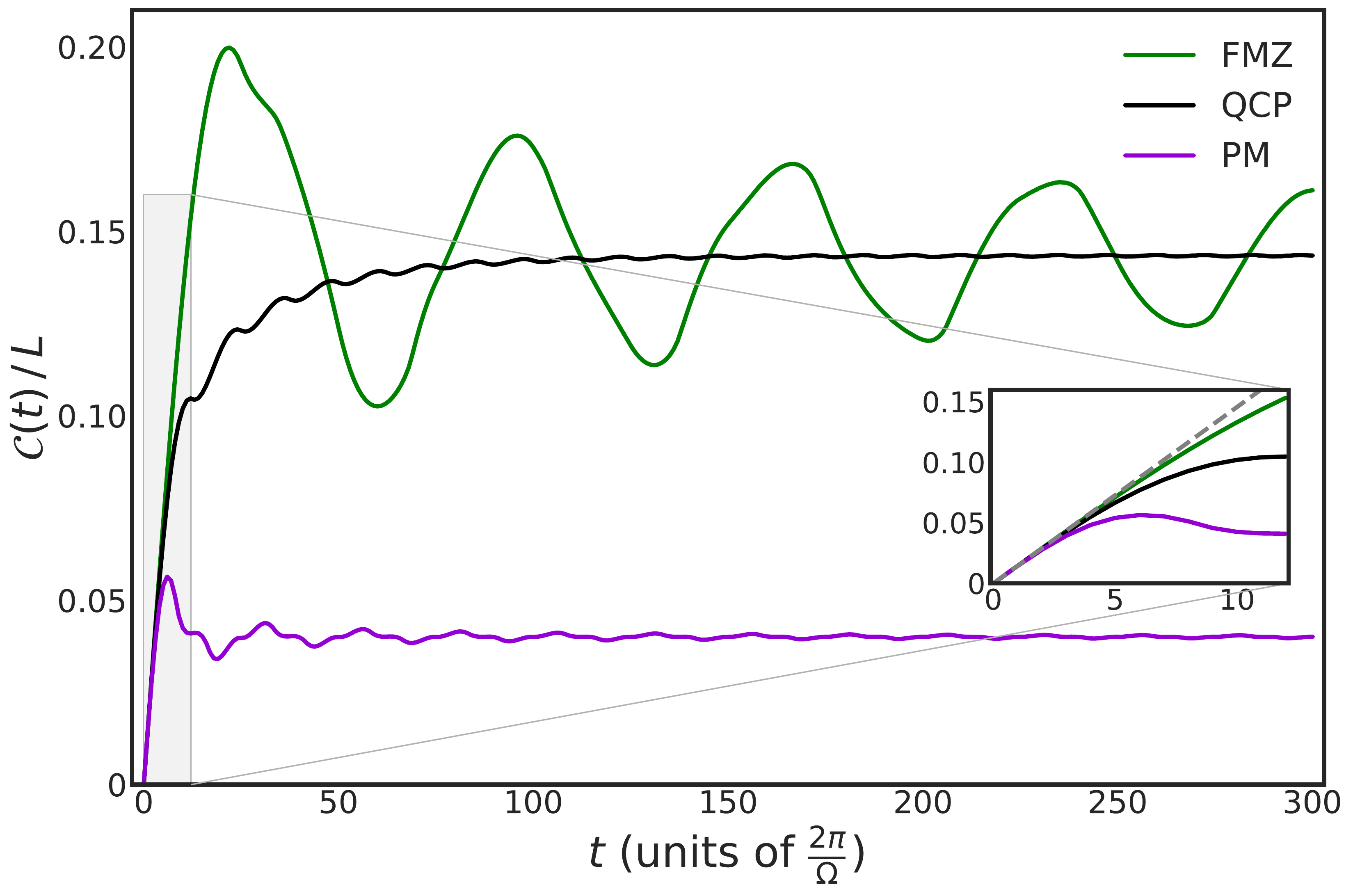}
    \caption{Time evolution of the complexity \eq{cpxt} near the Ising non-equilibrium transition. The parameters are $L=1000,\ell=2, J=0.01\Omega, g_1 = \Omega$ and varying 
    $\delta g_0^{(\ell)}=(0,J,2J)$, corresponding respectively to the ferromagnetic phase (FMZ), the quantum critical point (QCP), and the paramagnetic phase (PM). The dashed line in the inset shows the universal linear growth \eq{cpxt_early} at early times.}
    \label{fig:driven}
\end{figure}

The full time evolution of $\mathcal{C}(t)$ is depicted in Figure \fig{driven} for the Ising-like non-equilibrium QPT controlled by $g_0$. The early time behavior is readily obtained by a series expansion of \eq{cpxt}, with the summation over momenta performed analytically for the leading term to yield
\begin{equation}\label{eq:cpxt_early}
\mathcal{C}(t\to0) 
= \frac{2J\,\vert\gamma^{(\ell)}\vert}{\sin{\frac{\pi}{L}}}\,t+\mathcal{O}(t^3)\,.
\end{equation}
Note that in the thermodynamic limit $L\to\infty$ one has a volume law, ${\cal C}\sim L$. Interestingly, the linear growth at early times is independent of the constant field $g_0$. The inset in Figure \fig{driven} shows this universal early time behavior, which can be estimated to hold up to a time scale $t_*(g_0)\sim\min_k|2\delta g_0^{(\ell)}-\omega_k|^{-1}\approx\big|2|g_0- g_0^{(\ell)}|+2J\big|^{-1}$. 

The complexity clearly distinguishes between the two phases and the critical point -- in particular, it never equilibrates for $g_0$ in the FMZ phase. In the PM phase, it reaches the steady value ${\cal C}_{\infty}^{\rm PM}$ more rapidly for increasingly $g_0$, as one can infer from $t_{\ast}(g_0)$ estimated above and confirm numerically. Note that ${\cal C}_{\infty}^{\rm PM}$ is bounded from above by the value at the critical point ${\cal C}_{\infty}^{\rm QCP}$ and, in particular, it decreases as $g_0$ grows. Physically, this is an expression of the disordered character of the PM phase: complex (i.e., non-local) operations are required to create order in a state prepared on it, while simple (local) operations, like a phase shift, would maintain the disorder of such state. When $g_0$ is large, the effect of the driving field is suppressed and does not favor the possibility of creating operators complex enough to order the system, keeping it close to the initial paramagnetic ground state.

\begin{figure}
\centering
\subfloat{%
  \begin{minipage}{\columnwidth}
  \centering
  \subsubfloat{\includegraphics[scale=0.122,trim=10 0 10 0,clip]{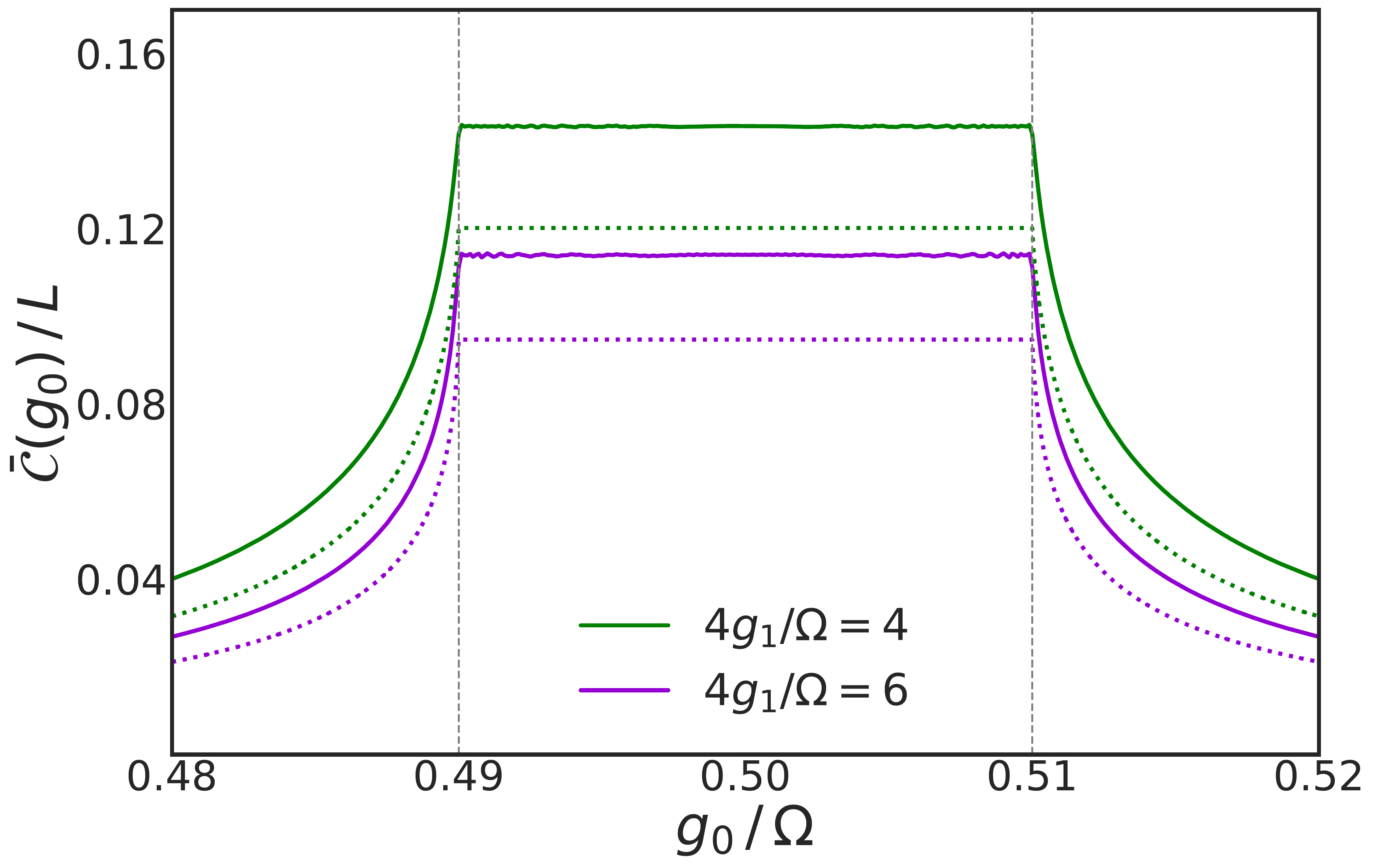}}{(a)}
  \quad
  \subsubfloat{\includegraphics[scale=0.122,trim=10 0 10 0,clip]{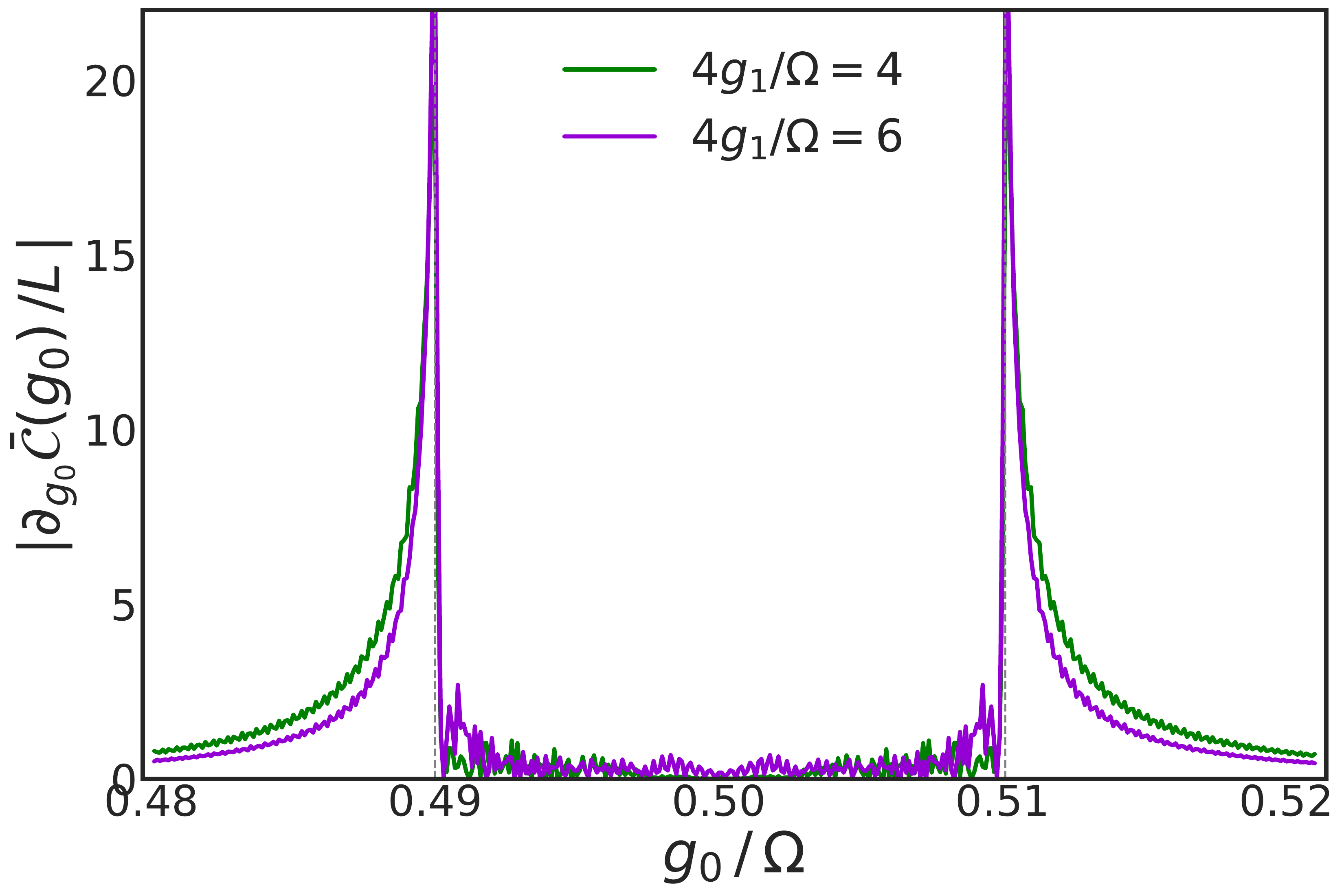}}{(b)}
  \subsubfloat{\includegraphics[scale=0.121,trim=10 0 10 0,clip]{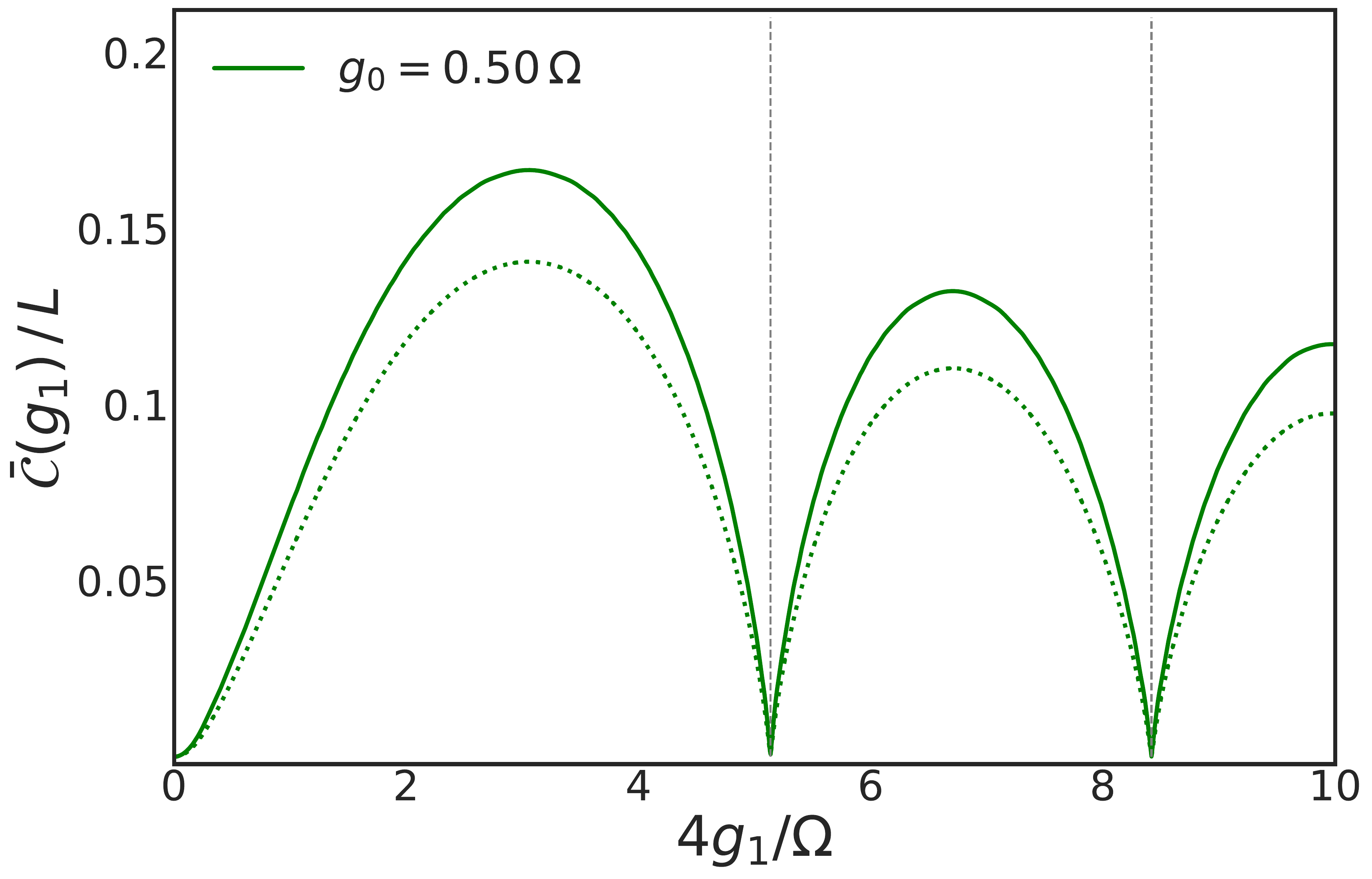}}{(c)}
  \quad
  \subsubfloat{\includegraphics[scale=0.119,trim=10 0 10 0,clip]{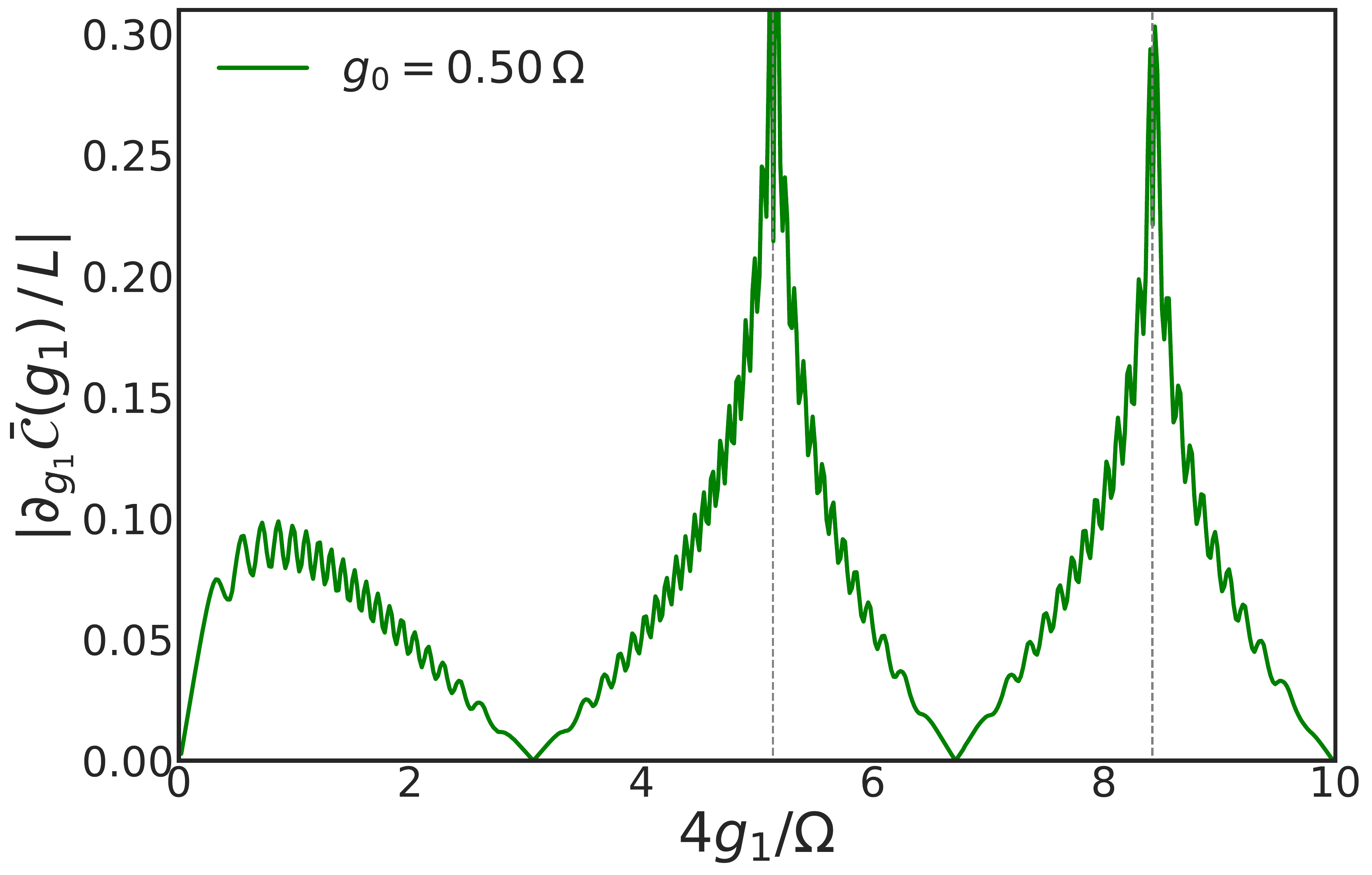}}{(d)}
  \end{minipage}}
\caption{Time-averaged complexity $\overline{\cal C}$ over $T=1000$ periods and its derivatives for $\ell=2, \Omega = \pi, J=0.01\Omega, L=1000$. (a) $\overline{\cal C}$ close to the Ising QPT for two values of $g_1$; the dotted lines show the corresponding Floquet mode complexity $\mathcal{C}^-$; (b) singular behavior of the first derivative of $\overline{\cal C}$ at the QPT points; (c) and (d) repeat the analysis of (a) and (b) for the first two anisotropic QPT points.}
\label{fig:Cbar}
\end{figure}

The critical behavior becomes more evident in terms of the time-averaged complexity 
\begin{equation}
\overline{{\cal C}} = \lim_{T\to\infty}\frac{1}{T}\int_{0}^T \dd t\, {\cal C}(t)\,.
\end{equation}
This quantity develops a non-analytic behavior at the QCP, as shown in Figure \fig{Cbar}. Such discontinuity becomes manifest as divergences in the derivatives at the critical points. This critical behavior is reminiscent from the behavior of the complexity in the undriven Ising model, which is discussed in Appendix \ref{app:Ising}.

Independently of $g_0$ it is evident that the complexity vanishes at the special anisotropic QPT points $\gamma^{(\ell)}=0$ designed by tuning $g_1$ and $\Omega$ to the Bessel zeros. This is a manifestation of the previously mentioned dynamic localization or CDT phenomenon happening at these points that freezes the quantum dynamics to the initial paramagnetic state. We also note that \eq{cpxt} is symmetric under $\gamma^{(\ell)}\to-\gamma^{(\ell)}$, showing that the complexity is unable to distinguish between the FMY and FMZ phases separated by the CDT point. Near these points, we can check that $\overline{{\cal C}}\propto \vert \gamma^{(\ell)}\vert\propto\vert g_1-g_1^c\vert$ to first order, which explains the type of non-analyticity observed in Figure \fig{Cbar}(c). Such behavior is essentially due to the complexity of the Floquet mode $\Phi^-$, since in this limit the Bogoliubov angle \eq{thetak} approaches zero and, therefore, the amplitude for positive mode in \eq{Psi_sol}, $A^+_{k,\ell} = - \sin{\vartheta_{k,\ell}}$, vanishes.

In fact, the similarity between $\overline{{\cal C}}$ and the complexity of Floquet modes is to be expected on more general grounds. At late times, after transients die out, the system synchronizes with the driving field and the dynamics is known to be governed by the Floquet modes \cite{Russomanno_2012}. 
Indeed, one can take a step further and make a concrete comparison by explicitly evaluating the complexity for each of 
the Floquet modes. 
We first note that those are easily put in the convenient form \eq{targetspinor},
\begin{align}\label{eq:targetspinor}
\Phi_{k}^{+(\ell)}(t) &\simeq
\begin{pmatrix}
e^{-\ii\,\alpha^{(\ell)}(t)-\ii\,\pi}\sin(\vartheta_{k,\ell}-\tfrac{\pi}{2})\\
\cos(\vartheta_{k,\ell}-\tfrac{\pi}{2})
\end{pmatrix}\notag\\
\Phi_{k}^{-(\ell)}(t) &\simeq 
\begin{pmatrix}
e^{-\ii\,\alpha^{(\ell)}(t)}\sin\vartheta_{k,\ell}\\
\cos\vartheta_{k,\ell}
\end{pmatrix}
,
\end{align}
from which the complexity follows trivially by paralleling the previous calculation and will be constant in time, namely $\mathcal{C}^+ = \sum_{k>0} \left|\vartheta_{k,\ell}-\tfrac{\pi}{2}\right|$ and $\mathcal{C}^- = \sum_{k>0} \left|\vartheta_{k,\ell}\right|$. 
When $t\to\infty$ we expect that the $e^{-2\ii\epsilon_{k,\ell}t}$ oscillations in \eq{uvsol} results in small contributions to the time-averaged complexity due to destructive interference (the same cannot be said about the $e^{-\ii\alpha^{(\ell)}(t)}$ prefactor, which contains the resonant term that survives to wild oscillations), so that the main contributions to $\overline{\mathcal{C}}$ come from the Floquet state $\Phi^{-(\ell)}_k(t)$. In other words, $\mathcal{C}(t\to\infty)\sim\mathcal{C}^-$ and, as consequence, the time average $\overline{\mathcal{C}}$ should replicate the behavior of $\mathcal{C}^-$, as indeed seen in Figure \fig{Cbar}(a) and (c).

\section{Final remarks}

We have studied the Floquet dynamics of Nielsen's circuit complexity for the Ising model driven by a time periodic transverse field. At high enough driving frequency, the model is analytically tractable and admits an exact determination of the non-equilibrium phase transitions induced by the external field. Here we showed that the complexity is able to diagnose these non-equilibrium QPTs, extending previous ideas in the literature for quantum quench protocols and hence strengthening the case for complexity as a tool to understand the non-equilibrium physics of many-body systems. In particular, we showed that for a paramagnetic reference state, the complexity of the instantaneous time-evolved state can only equilibrate at large times provided the critical point is not crossed, otherwise it oscillates indefinitely in time. We also proved that the early time transient behavior of the complexity is linear and independent of the constant driving field $g_0$ up to a time scale inversely proportional to $g_0$. The long-time average of the complexity presents non-analytical behavior at the critical points, which can be traced back to the fact that the asymptotic dynamics is governed by the Floquet modes. 

The sensitivity of the circuit complexity to non-equilibrium critical phenomena encourages us to investigate its role in the description of dynamical phase transitions \cite{Heyl_2018}, which are characterized by a non-analytical behavior in the time domain and whose scaling and universality properties are not fully understood. These phenomena can be engineered using quantum quenches or in periodically driven systems similar to the one studied here \cite{Yang_2019}. 
The time evolution of complexity (analogue of Figure \fig{driven}) should develop a singular behavior at the critical time $t_c$ and may help in the classification of non-trivial topological Floquet phases. This is work in progress. 

Another interesting future direction to pursue would be to see how the present analysis generalizes to the case of interacting models, where more elaborate gates than simple $SU(2)$ rotations used here are required to produce physically interesting states. Here the set of integrable spin chains immediately comes to mind \cite{Gritsev_2017}.
A more ambitious goal would be the study of a many-body localization/thermal transition, which can be modeled with a Floquet system with no conserved charges \cite{Zhang_2016}.

\begin{acknowledgments}
We are grateful to Sebas \"Eliens and Diego Trancanelli for discussions and comments. G.C. thanks financial support from MEC and MCTIC. Work of D. T. supported by FAPESP grant 2017/02300-2. 
\end{acknowledgments}

\appendix

\section{Complexity in the Ising model}\label{app:Ising}

\begin{figure}[b]
\centering
\subfloat{
  \begin{minipage}{\columnwidth}
  \centering
  \subsubfloat{\includegraphics[width=0.48\columnwidth]{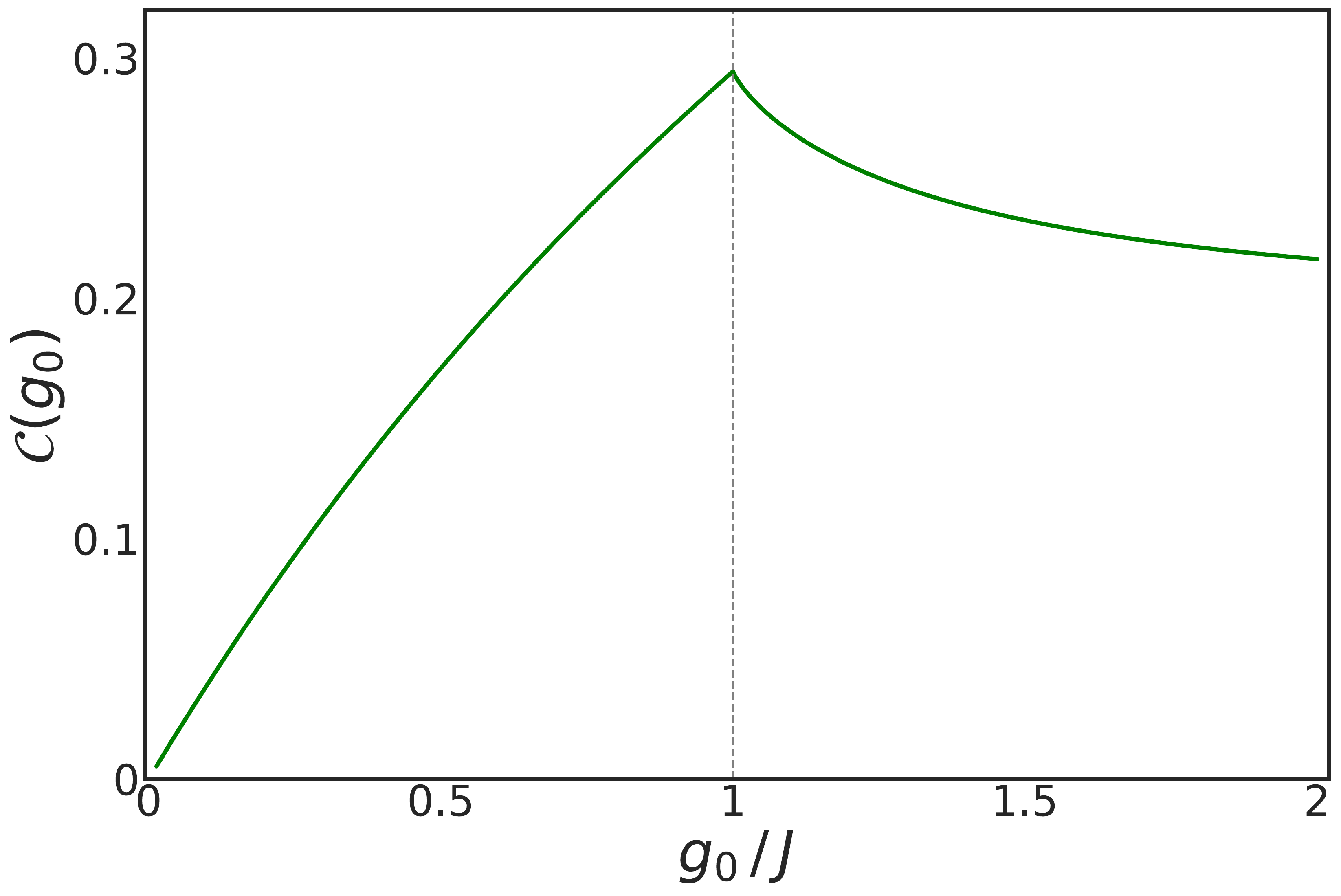}}{(a)}
  \quad
  \subsubfloat{\includegraphics[width=0.48\columnwidth]{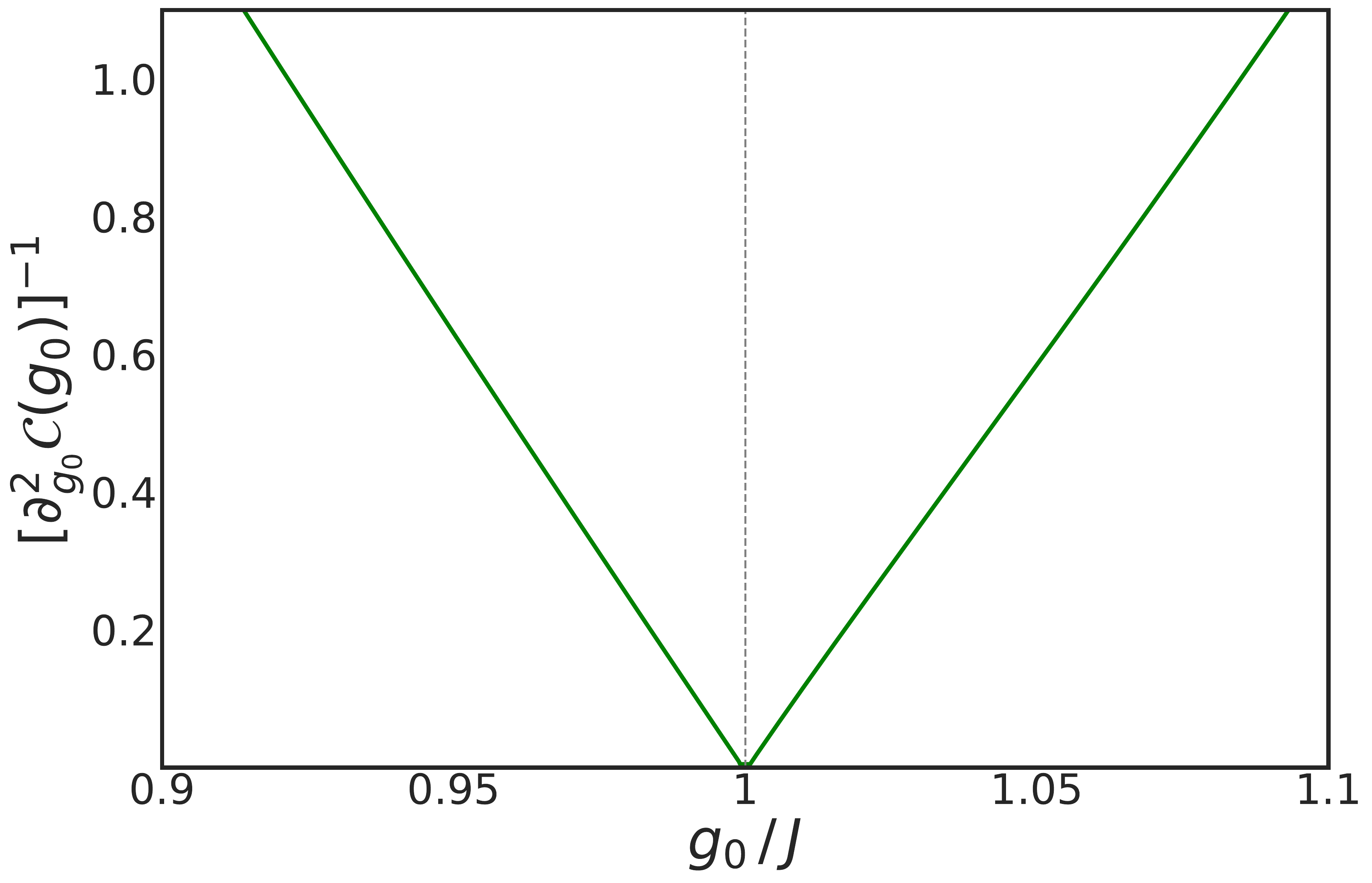}}{(b)}
  \end{minipage}}
\caption{(a) Complexity of the ground state of the undriven Ising model with $J=1$. (b) $\vert g_0-J\vert^{-1}$ behavior of the second derivative near the critical point.}
\label{fig:Ising_complexity}
\end{figure}

In order to further illustrate how the circuit complexity can be used to diagnose an equilibrium QPT as well, let us evaluate it for the standard Ising model with a constant transverse field. We take both reference and target states belonging to the ground state manifold, that is, they can be written as $(\cos\eta_k^{\rm (R,T)}+\ii\sin\eta_k^{\rm (R,T)})^{\otimes k>0}\vert 0\rangle$ such that the complexity assumes the simple form ${\cal C} = \sum_k\vert\Delta\eta_k\vert$, where $\Delta\eta_k$ is the relative Bogoliubov angle between $\vert R\rangle$ and $\vert T\rangle$. Here it is straightforward to work even in the infinite chain limit, where 
\begin{equation}\label{eq:Ising_complexity}
    {\cal C} = \frac{1}{2\pi} \int_0^{\pi}\dd k \,\vert\Delta\eta_k\vert\, .
\end{equation}
Using the usual spectrum and Bogoliubov angle of the Ising model, one can easily compute this object which is illustrated in Figure \ref{fig:Ising_complexity} where, for simplicity, we have chosen $\eta_k^{\rm (R)} = 0$. The first derivative is discontinuous at the quantum critical point, $g_0 = J$, while the second-derivative diverges with a unit critical exponent, that is $\sim \vert g_0 -J\vert^{-1}$, as shown in Figure \ref{fig:Ising_complexity}(b).

\bibliography{ComplexityFloquet_final}

\providecommand{\noopsort}[1]{}\providecommand{\singleletter}[1]{#1}%
\begin{thebibliography}{48}%
\makeatletter
\providecommand \@ifxundefined [1]{%
 \@ifx{#1\undefined}
}%
\providecommand \@ifnum [1]{%
 \ifnum #1\expandafter \@firstoftwo
 \else \expandafter \@secondoftwo
 \fi
}%
\providecommand \@ifx [1]{%
 \ifx #1\expandafter \@firstoftwo
 \else \expandafter \@secondoftwo
 \fi
}%
\providecommand \natexlab [1]{#1}%
\providecommand \enquote  [1]{``#1''}%
\providecommand \bibnamefont  [1]{#1}%
\providecommand \bibfnamefont [1]{#1}%
\providecommand \citenamefont [1]{#1}%
\providecommand \href@noop [0]{\@secondoftwo}%
\providecommand \href [0]{\begingroup \@sanitize@url \@href}%
\providecommand \@href[1]{\@@startlink{#1}\@@href}%
\providecommand \@@href[1]{\endgroup#1\@@endlink}%
\providecommand \@sanitize@url [0]{\catcode `\\12\catcode `\$12\catcode
  `\&12\catcode `\#12\catcode `\^12\catcode `\_12\catcode `\%12\relax}%
\providecommand \@@startlink[1]{}%
\providecommand \@@endlink[0]{}%
\providecommand \url  [0]{\begingroup\@sanitize@url \@url }%
\providecommand \@url [1]{\endgroup\@href {#1}{\urlprefix }}%
\providecommand \urlprefix  [0]{URL }%
\providecommand \Eprint [0]{\href }%
\providecommand \doibase [0]{https://doi.org/}%
\providecommand \selectlanguage [0]{\@gobble}%
\providecommand \bibinfo  [0]{\@secondoftwo}%
\providecommand \bibfield  [0]{\@secondoftwo}%
\providecommand \translation [1]{[#1]}%
\providecommand \BibitemOpen [0]{}%
\providecommand \bibitemStop [0]{}%
\providecommand \bibitemNoStop [0]{.\EOS\space}%
\providecommand \EOS [0]{\spacefactor3000\relax}%
\providecommand \BibitemShut  [1]{\csname bibitem#1\endcsname}%
\let\auto@bib@innerbib\@empty
\bibitem [{\citenamefont {Kibble}(1976)}]{Kibble:1976sj}%
  \BibitemOpen
  \bibfield  {author} {\bibinfo {author} {\bibfnamefont {T.}~\bibnamefont
  {Kibble}},\ }\bibfield  {title} {\bibinfo {title} {{Topology of Cosmic
  Domains and Strings}},\ }\href {https://doi.org/10.1088/0305-4470/9/8/029}
  {\bibfield  {journal} {\bibinfo  {journal} {J. Phys. A}\ }\textbf {\bibinfo
  {volume} {9}},\ \bibinfo {pages} {1387} (\bibinfo {year} {1976})}\BibitemShut
  {NoStop}%
\bibitem [{\citenamefont {Zurek}(1985)}]{Zurek:1985qw}%
  \BibitemOpen
  \bibfield  {author} {\bibinfo {author} {\bibfnamefont {W.}~\bibnamefont
  {Zurek}},\ }\bibfield  {title} {\bibinfo {title} {{Cosmological Experiments
  in Superfluid Helium?}},\ }\href {https://doi.org/10.1038/317505a0}
  {\bibfield  {journal} {\bibinfo  {journal} {Nature}\ }\textbf {\bibinfo
  {volume} {317}},\ \bibinfo {pages} {505} (\bibinfo {year}
  {1985})}\BibitemShut {NoStop}%
\bibitem [{\citenamefont
  {Dziarmaga}(2005{\natexlab{a}})}]{PhysRevLett.95.245701}%
  \BibitemOpen
  \bibfield  {author} {\bibinfo {author} {\bibfnamefont {J.}~\bibnamefont
  {Dziarmaga}},\ }\bibfield  {title} {\bibinfo {title} {Dynamics of a quantum
  phase transition: Exact solution of the quantum ising model},\ }\href
  {https://doi.org/10.1103/PhysRevLett.95.245701} {\bibfield  {journal}
  {\bibinfo  {journal} {Phys. Rev. Lett.}\ }\textbf {\bibinfo {volume} {95}},\
  \bibinfo {pages} {245701} (\bibinfo {year} {2005}{\natexlab{a}})}\BibitemShut
  {NoStop}%
\bibitem [{\citenamefont {Polkovnikov}(2005)}]{PhysRevB.72.161201}%
  \BibitemOpen
  \bibfield  {author} {\bibinfo {author} {\bibfnamefont {A.}~\bibnamefont
  {Polkovnikov}},\ }\bibfield  {title} {\bibinfo {title} {Universal adiabatic
  dynamics in the vicinity of a quantum critical point},\ }\href
  {https://doi.org/10.1103/PhysRevB.72.161201} {\bibfield  {journal} {\bibinfo
  {journal} {Phys. Rev. B}\ }\textbf {\bibinfo {volume} {72}},\ \bibinfo
  {pages} {161201} (\bibinfo {year} {2005})}\BibitemShut {NoStop}%
\bibitem [{\citenamefont {Zurek}\ \emph {et~al.}(2005)\citenamefont {Zurek},
  \citenamefont {Dorner},\ and\ \citenamefont
  {Zoller}}]{PhysRevLett.95.105701}%
  \BibitemOpen
  \bibfield  {author} {\bibinfo {author} {\bibfnamefont {W.~H.}\ \bibnamefont
  {Zurek}}, \bibinfo {author} {\bibfnamefont {U.}~\bibnamefont {Dorner}},\ and\
  \bibinfo {author} {\bibfnamefont {P.}~\bibnamefont {Zoller}},\ }\bibfield
  {title} {\bibinfo {title} {Dynamics of a quantum phase transition},\ }\href
  {https://doi.org/10.1103/PhysRevLett.95.105701} {\bibfield  {journal}
  {\bibinfo  {journal} {Phys. Rev. Lett.}\ }\textbf {\bibinfo {volume} {95}},\
  \bibinfo {pages} {105701} (\bibinfo {year} {2005})}\BibitemShut {NoStop}%
\bibitem [{\citenamefont {GU}(2010)}]{GU_2010}%
  \BibitemOpen
  \bibfield  {author} {\bibinfo {author} {\bibfnamefont {S.-J.}\ \bibnamefont
  {GU}},\ }\bibfield  {title} {\bibinfo {title} {Fidelity approach to quantum
  phase transitions},\ }\href {https://doi.org/10.1142/s0217979210056335}
  {\bibfield  {journal} {\bibinfo  {journal} {International Journal of Modern
  Physics B}\ }\textbf {\bibinfo {volume} {24}},\ \bibinfo {pages}
  {4371–4458} (\bibinfo {year} {2010})}\BibitemShut {NoStop}%
\bibitem [{\citenamefont {Osterloh}\ \emph {et~al.}(2002)\citenamefont
  {Osterloh}, \citenamefont {Amico}, \citenamefont {Falci},\ and\ \citenamefont
  {Fazio}}]{Osterloh_2002}%
  \BibitemOpen
  \bibfield  {author} {\bibinfo {author} {\bibfnamefont {A.}~\bibnamefont
  {Osterloh}}, \bibinfo {author} {\bibfnamefont {L.}~\bibnamefont {Amico}},
  \bibinfo {author} {\bibfnamefont {G.}~\bibnamefont {Falci}},\ and\ \bibinfo
  {author} {\bibfnamefont {R.}~\bibnamefont {Fazio}},\ }\bibfield  {title}
  {\bibinfo {title} {Scaling of entanglement close to a quantum phase
  transition},\ }\href {https://doi.org/10.1038/416608a} {\bibfield  {journal}
  {\bibinfo  {journal} {Nature}\ }\textbf {\bibinfo {volume} {416}},\ \bibinfo
  {pages} {608–610} (\bibinfo {year} {2002})}\BibitemShut {NoStop}%
\bibitem [{\citenamefont {Vidal}\ \emph {et~al.}(2003)\citenamefont {Vidal},
  \citenamefont {Latorre}, \citenamefont {Rico},\ and\ \citenamefont
  {Kitaev}}]{Vidal_2003}%
  \BibitemOpen
  \bibfield  {author} {\bibinfo {author} {\bibfnamefont {G.}~\bibnamefont
  {Vidal}}, \bibinfo {author} {\bibfnamefont {J.~I.}\ \bibnamefont {Latorre}},
  \bibinfo {author} {\bibfnamefont {E.}~\bibnamefont {Rico}},\ and\ \bibinfo
  {author} {\bibfnamefont {A.}~\bibnamefont {Kitaev}},\ }\bibfield  {title}
  {\bibinfo {title} {Entanglement in quantum critical phenomena},\ }\bibfield
  {journal} {\bibinfo  {journal} {Physical Review Letters}\ }\textbf {\bibinfo
  {volume} {90}},\ \href {https://doi.org/10.1103/physrevlett.90.227902}
  {10.1103/physrevlett.90.227902} (\bibinfo {year} {2003})\BibitemShut
  {NoStop}%
\bibitem [{\citenamefont {Zeng}\ \emph {et~al.}(2015)\citenamefont {Zeng},
  \citenamefont {Chen}, \citenamefont {Zhou},\ and\ \citenamefont
  {Wen}}]{zeng2015quantum}%
  \BibitemOpen
  \bibfield  {author} {\bibinfo {author} {\bibfnamefont {B.}~\bibnamefont
  {Zeng}}, \bibinfo {author} {\bibfnamefont {X.}~\bibnamefont {Chen}}, \bibinfo
  {author} {\bibfnamefont {D.-L.}\ \bibnamefont {Zhou}},\ and\ \bibinfo
  {author} {\bibfnamefont {X.-G.}\ \bibnamefont {Wen}},\ }\href@noop {}
  {\bibinfo {title} {Quantum information meets quantum matter -- from quantum
  entanglement to topological phase in many-body systems}} (\bibinfo {year}
  {2015}),\ \Eprint {https://arxiv.org/abs/1508.02595} {arXiv:1508.02595
  [cond-mat.str-el]} \BibitemShut {NoStop}%
\bibitem [{\citenamefont {Liu}\ \emph {et~al.}(2020)\citenamefont {Liu},
  \citenamefont {Whitsitt}, \citenamefont {Curtis}, \citenamefont {Lundgren},
  \citenamefont {Titum}, \citenamefont {Yang}, \citenamefont {Garrison},\ and\
  \citenamefont {Gorshkov}}]{Liu_2020}%
  \BibitemOpen
  \bibfield  {author} {\bibinfo {author} {\bibfnamefont {F.}~\bibnamefont
  {Liu}}, \bibinfo {author} {\bibfnamefont {S.}~\bibnamefont {Whitsitt}},
  \bibinfo {author} {\bibfnamefont {J.~B.}\ \bibnamefont {Curtis}}, \bibinfo
  {author} {\bibfnamefont {R.}~\bibnamefont {Lundgren}}, \bibinfo {author}
  {\bibfnamefont {P.}~\bibnamefont {Titum}}, \bibinfo {author} {\bibfnamefont
  {Z.-C.}\ \bibnamefont {Yang}}, \bibinfo {author} {\bibfnamefont {J.~R.}\
  \bibnamefont {Garrison}},\ and\ \bibinfo {author} {\bibfnamefont {A.~V.}\
  \bibnamefont {Gorshkov}},\ }\bibfield  {title} {\bibinfo {title} {Circuit
  complexity across a topological phase transition},\ }\bibfield  {journal}
  {\bibinfo  {journal} {Physical Review Research}\ }\textbf {\bibinfo {volume}
  {2}},\ \href {https://doi.org/10.1103/physrevresearch.2.013323}
  {10.1103/physrevresearch.2.013323} (\bibinfo {year} {2020})\BibitemShut
  {NoStop}%
\bibitem [{\citenamefont {Xiong}\ \emph {et~al.}(2020)\citenamefont {Xiong},
  \citenamefont {Yao},\ and\ \citenamefont {Yan}}]{PhysRevB.101.174305}%
  \BibitemOpen
  \bibfield  {author} {\bibinfo {author} {\bibfnamefont {Z.}~\bibnamefont
  {Xiong}}, \bibinfo {author} {\bibfnamefont {D.-X.}\ \bibnamefont {Yao}},\
  and\ \bibinfo {author} {\bibfnamefont {Z.}~\bibnamefont {Yan}},\ }\bibfield
  {title} {\bibinfo {title} {Nonanalyticity of circuit complexity across
  topological phase transitions},\ }\href
  {https://doi.org/10.1103/PhysRevB.101.174305} {\bibfield  {journal} {\bibinfo
   {journal} {Phys. Rev. B}\ }\textbf {\bibinfo {volume} {101}},\ \bibinfo
  {pages} {174305} (\bibinfo {year} {2020})}\BibitemShut {NoStop}%
\bibitem [{\citenamefont {Jaiswal}\ \emph {et~al.}(2020)\citenamefont
  {Jaiswal}, \citenamefont {Gautam},\ and\ \citenamefont
  {Sarkar}}]{jaiswal2020complexity}%
  \BibitemOpen
  \bibfield  {author} {\bibinfo {author} {\bibfnamefont {N.}~\bibnamefont
  {Jaiswal}}, \bibinfo {author} {\bibfnamefont {M.}~\bibnamefont {Gautam}},\
  and\ \bibinfo {author} {\bibfnamefont {T.}~\bibnamefont {Sarkar}},\
  }\href@noop {} {\bibinfo {title} {Complexity and information geometry in spin
  chains}} (\bibinfo {year} {2020}),\ \Eprint
  {https://arxiv.org/abs/2005.03532} {arXiv:2005.03532 [cond-mat.stat-mech]}
  \BibitemShut {NoStop}%
\bibitem [{\citenamefont {Nielsen}(2005)}]{nielsen2005geometric}%
  \BibitemOpen
  \bibfield  {author} {\bibinfo {author} {\bibfnamefont {M.~A.}\ \bibnamefont
  {Nielsen}},\ }\href@noop {} {\bibinfo {title} {A geometric approach to
  quantum circuit lower bounds}} (\bibinfo {year} {2005}),\ \Eprint
  {https://arxiv.org/abs/quant-ph/0502070} {arXiv:quant-ph/0502070 [quant-ph]}
  \BibitemShut {NoStop}%
\bibitem [{\citenamefont {Nielsen}(2006)}]{Nielsen_2006}%
  \BibitemOpen
  \bibfield  {author} {\bibinfo {author} {\bibfnamefont {M.~A.}\ \bibnamefont
  {Nielsen}},\ }\bibfield  {title} {\bibinfo {title} {Quantum computation as
  geometry},\ }\href {https://doi.org/10.1126/science.1121541} {\bibfield
  {journal} {\bibinfo  {journal} {Science}\ }\textbf {\bibinfo {volume}
  {311}},\ \bibinfo {pages} {1133–1135} (\bibinfo {year} {2006})}\BibitemShut
  {NoStop}%
\bibitem [{\citenamefont {Stanford}\ and\ \citenamefont
  {Susskind}(2014)}]{Stanford:2014jda}%
  \BibitemOpen
  \bibfield  {author} {\bibinfo {author} {\bibfnamefont {D.}~\bibnamefont
  {Stanford}}\ and\ \bibinfo {author} {\bibfnamefont {L.}~\bibnamefont
  {Susskind}},\ }\bibfield  {title} {\bibinfo {title} {{Complexity and Shock
  Wave Geometries}},\ }\href {https://doi.org/10.1103/PhysRevD.90.126007}
  {\bibfield  {journal} {\bibinfo  {journal} {Phys. Rev. D}\ }\textbf {\bibinfo
  {volume} {90}},\ \bibinfo {pages} {126007} (\bibinfo {year} {2014})},\
  \Eprint {https://arxiv.org/abs/1406.2678} {arXiv:1406.2678 [hep-th]}
  \BibitemShut {NoStop}%
\bibitem [{\citenamefont {Brown}\ \emph {et~al.}(2016)\citenamefont {Brown},
  \citenamefont {Roberts}, \citenamefont {Susskind}, \citenamefont {Swingle},\
  and\ \citenamefont {Zhao}}]{Brown:2015bva}%
  \BibitemOpen
  \bibfield  {author} {\bibinfo {author} {\bibfnamefont {A.~R.}\ \bibnamefont
  {Brown}}, \bibinfo {author} {\bibfnamefont {D.~A.}\ \bibnamefont {Roberts}},
  \bibinfo {author} {\bibfnamefont {L.}~\bibnamefont {Susskind}}, \bibinfo
  {author} {\bibfnamefont {B.}~\bibnamefont {Swingle}},\ and\ \bibinfo {author}
  {\bibfnamefont {Y.}~\bibnamefont {Zhao}},\ }\bibfield  {title} {\bibinfo
  {title} {{Holographic Complexity Equals Bulk Action?}},\ }\href
  {https://doi.org/10.1103/PhysRevLett.116.191301} {\bibfield  {journal}
  {\bibinfo  {journal} {Phys. Rev. Lett.}\ }\textbf {\bibinfo {volume} {116}},\
  \bibinfo {pages} {191301} (\bibinfo {year} {2016})},\ \Eprint
  {https://arxiv.org/abs/1509.07876} {arXiv:1509.07876 [hep-th]} \BibitemShut
  {NoStop}%
\bibitem [{\citenamefont {Jefferson}\ and\ \citenamefont
  {Myers}(2017)}]{Jefferson:2017sdb}%
  \BibitemOpen
  \bibfield  {author} {\bibinfo {author} {\bibfnamefont {R.}~\bibnamefont
  {Jefferson}}\ and\ \bibinfo {author} {\bibfnamefont {R.~C.}\ \bibnamefont
  {Myers}},\ }\bibfield  {title} {\bibinfo {title} {{Circuit complexity in
  quantum field theory}},\ }\href {https://doi.org/10.1007/JHEP10(2017)107}
  {\bibfield  {journal} {\bibinfo  {journal} {JHEP}\ }\textbf {\bibinfo
  {volume} {10}},\ \bibinfo {pages} {107}},\ \Eprint
  {https://arxiv.org/abs/1707.08570} {arXiv:1707.08570 [hep-th]} \BibitemShut
  {NoStop}%
\bibitem [{\citenamefont {Caputa}\ and\ \citenamefont
  {Magan}(2019)}]{Caputa:2018kdj}%
  \BibitemOpen
  \bibfield  {author} {\bibinfo {author} {\bibfnamefont {P.}~\bibnamefont
  {Caputa}}\ and\ \bibinfo {author} {\bibfnamefont {J.~M.}\ \bibnamefont
  {Magan}},\ }\bibfield  {title} {\bibinfo {title} {{Quantum Computation as
  Gravity}},\ }\href {https://doi.org/10.1103/PhysRevLett.122.231302}
  {\bibfield  {journal} {\bibinfo  {journal} {Phys. Rev. Lett.}\ }\textbf
  {\bibinfo {volume} {122}},\ \bibinfo {pages} {231302} (\bibinfo {year}
  {2019})},\ \Eprint {https://arxiv.org/abs/1807.04422} {arXiv:1807.04422
  [hep-th]} \BibitemShut {NoStop}%
\bibitem [{\citenamefont {Balasubramanian}\ \emph {et~al.}(2020)\citenamefont
  {Balasubramanian}, \citenamefont {Decross}, \citenamefont {Kar},\ and\
  \citenamefont {Parrikar}}]{Balasubramanian:2019wgd}%
  \BibitemOpen
  \bibfield  {author} {\bibinfo {author} {\bibfnamefont {V.}~\bibnamefont
  {Balasubramanian}}, \bibinfo {author} {\bibfnamefont {M.}~\bibnamefont
  {Decross}}, \bibinfo {author} {\bibfnamefont {A.}~\bibnamefont {Kar}},\ and\
  \bibinfo {author} {\bibfnamefont {O.}~\bibnamefont {Parrikar}},\ }\bibfield
  {title} {\bibinfo {title} {{Quantum Complexity of Time Evolution with Chaotic
  Hamiltonians}},\ }\href {https://doi.org/10.1007/JHEP01(2020)134} {\bibfield
  {journal} {\bibinfo  {journal} {JHEP}\ }\textbf {\bibinfo {volume} {01}},\
  \bibinfo {pages} {134}},\ \Eprint {https://arxiv.org/abs/1905.05765}
  {arXiv:1905.05765 [hep-th]} \BibitemShut {NoStop}%
\bibitem [{\citenamefont {Polkovnikov}\ \emph {et~al.}(2011)\citenamefont
  {Polkovnikov}, \citenamefont {Sengupta}, \citenamefont {Silva},\ and\
  \citenamefont {Vengalattore}}]{Polkovnikov_2011}%
  \BibitemOpen
  \bibfield  {author} {\bibinfo {author} {\bibfnamefont {A.}~\bibnamefont
  {Polkovnikov}}, \bibinfo {author} {\bibfnamefont {K.}~\bibnamefont
  {Sengupta}}, \bibinfo {author} {\bibfnamefont {A.}~\bibnamefont {Silva}},\
  and\ \bibinfo {author} {\bibfnamefont {M.}~\bibnamefont {Vengalattore}},\
  }\bibfield  {title} {\bibinfo {title} {Colloquium: Nonequilibrium dynamics of
  closed interacting quantum systems},\ }\href
  {https://doi.org/10.1103/revmodphys.83.863} {\bibfield  {journal} {\bibinfo
  {journal} {Reviews of Modern Physics}\ }\textbf {\bibinfo {volume} {83}},\
  \bibinfo {pages} {863–883} (\bibinfo {year} {2011})}\BibitemShut {NoStop}%
\bibitem [{\citenamefont {Mitra}(2018)}]{Mitra_2018}%
  \BibitemOpen
  \bibfield  {author} {\bibinfo {author} {\bibfnamefont {A.}~\bibnamefont
  {Mitra}},\ }\bibfield  {title} {\bibinfo {title} {Quantum quench dynamics},\
  }\href {https://doi.org/10.1146/annurev-conmatphys-031016-025451} {\bibfield
  {journal} {\bibinfo  {journal} {Annual Review of Condensed Matter Physics}\
  }\textbf {\bibinfo {volume} {9}},\ \bibinfo {pages} {245–259} (\bibinfo
  {year} {2018})}\BibitemShut {NoStop}%
\bibitem [{\citenamefont {Alves}\ and\ \citenamefont
  {Camilo}(2018)}]{Alves:2018qfv}%
  \BibitemOpen
  \bibfield  {author} {\bibinfo {author} {\bibfnamefont {D.~W.}\ \bibnamefont
  {Alves}}\ and\ \bibinfo {author} {\bibfnamefont {G.}~\bibnamefont {Camilo}},\
  }\bibfield  {title} {\bibinfo {title} {{Evolution of complexity following a
  quantum quench in free field theory}},\ }\href
  {https://doi.org/10.1007/JHEP06(2018)029} {\bibfield  {journal} {\bibinfo
  {journal} {JHEP}\ }\textbf {\bibinfo {volume} {06}},\ \bibinfo {pages}
  {029}},\ \Eprint {https://arxiv.org/abs/1804.00107} {arXiv:1804.00107
  [hep-th]} \BibitemShut {NoStop}%
\bibitem [{\citenamefont {Camargo}\ \emph {et~al.}(2019)\citenamefont
  {Camargo}, \citenamefont {Caputa}, \citenamefont {Das}, \citenamefont
  {Heller},\ and\ \citenamefont {Jefferson}}]{Camargo:2018eof}%
  \BibitemOpen
  \bibfield  {author} {\bibinfo {author} {\bibfnamefont {H.~A.}\ \bibnamefont
  {Camargo}}, \bibinfo {author} {\bibfnamefont {P.}~\bibnamefont {Caputa}},
  \bibinfo {author} {\bibfnamefont {D.}~\bibnamefont {Das}}, \bibinfo {author}
  {\bibfnamefont {M.~P.}\ \bibnamefont {Heller}},\ and\ \bibinfo {author}
  {\bibfnamefont {R.}~\bibnamefont {Jefferson}},\ }\bibfield  {title} {\bibinfo
  {title} {{Complexity as a novel probe of quantum quenches: universal scalings
  and purifications}},\ }\href {https://doi.org/10.1103/PhysRevLett.122.081601}
  {\bibfield  {journal} {\bibinfo  {journal} {Phys. Rev. Lett.}\ }\textbf
  {\bibinfo {volume} {122}},\ \bibinfo {pages} {081601} (\bibinfo {year}
  {2019})},\ \Eprint {https://arxiv.org/abs/1807.07075} {arXiv:1807.07075
  [hep-th]} \BibitemShut {NoStop}%
\bibitem [{\citenamefont {Liu}(2019)}]{Liu:2019qyx}%
  \BibitemOpen
  \bibfield  {author} {\bibinfo {author} {\bibfnamefont {S.}~\bibnamefont
  {Liu}},\ }\bibfield  {title} {\bibinfo {title} {{Complexity and scaling in
  quantum quench in $1 + 1$ dimensional fermionic field theories}},\ }\href
  {https://doi.org/10.1007/JHEP07(2019)104} {\bibfield  {journal} {\bibinfo
  {journal} {JHEP}\ }\textbf {\bibinfo {volume} {07}},\ \bibinfo {pages}
  {104}},\ \Eprint {https://arxiv.org/abs/1902.02945} {arXiv:1902.02945
  [hep-th]} \BibitemShut {NoStop}%
\bibitem [{\citenamefont {Ali}\ \emph {et~al.}(2018)\citenamefont {Ali},
  \citenamefont {Bhattacharyya}, \citenamefont {Haque}, \citenamefont {Kim},\
  and\ \citenamefont {Moynihan}}]{ali2018postquench}%
  \BibitemOpen
  \bibfield  {author} {\bibinfo {author} {\bibfnamefont {T.}~\bibnamefont
  {Ali}}, \bibinfo {author} {\bibfnamefont {A.}~\bibnamefont {Bhattacharyya}},
  \bibinfo {author} {\bibfnamefont {S.~S.}\ \bibnamefont {Haque}}, \bibinfo
  {author} {\bibfnamefont {E.~H.}\ \bibnamefont {Kim}},\ and\ \bibinfo {author}
  {\bibfnamefont {N.}~\bibnamefont {Moynihan}},\ }\href@noop {} {\bibinfo
  {title} {Post-quench evolution of distance and uncertainty in a topological
  system: Complexity, entanglement and revivals}} (\bibinfo {year} {2018}),\
  \Eprint {https://arxiv.org/abs/1811.05985} {arXiv:1811.05985 [hep-th]}
  \BibitemShut {NoStop}%
\bibitem [{\citenamefont {Eckardt}(2017)}]{Eckardt_2017}%
  \BibitemOpen
  \bibfield  {author} {\bibinfo {author} {\bibfnamefont {A.}~\bibnamefont
  {Eckardt}},\ }\bibfield  {title} {\bibinfo {title} {Colloquium: Atomic
  quantum gases in periodically driven optical lattices},\ }\bibfield
  {journal} {\bibinfo  {journal} {Reviews of Modern Physics}\ }\textbf
  {\bibinfo {volume} {89}},\ \href
  {https://doi.org/10.1103/revmodphys.89.011004} {10.1103/revmodphys.89.011004}
  (\bibinfo {year} {2017})\BibitemShut {NoStop}%
\bibitem [{\citenamefont {Oka}\ and\ \citenamefont
  {Kitamura}(2019)}]{Oka_2019}%
  \BibitemOpen
  \bibfield  {author} {\bibinfo {author} {\bibfnamefont {T.}~\bibnamefont
  {Oka}}\ and\ \bibinfo {author} {\bibfnamefont {S.}~\bibnamefont {Kitamura}},\
  }\bibfield  {title} {\bibinfo {title} {Floquet engineering of quantum
  materials},\ }\href
  {https://doi.org/10.1146/annurev-conmatphys-031218-013423} {\bibfield
  {journal} {\bibinfo  {journal} {Annual Review of Condensed Matter Physics}\
  }\textbf {\bibinfo {volume} {10}},\ \bibinfo {pages} {387–408} (\bibinfo
  {year} {2019})}\BibitemShut {NoStop}%
\bibitem [{\citenamefont {Lignier}\ \emph {et~al.}(2007)\citenamefont
  {Lignier}, \citenamefont {Sias}, \citenamefont {Ciampini}, \citenamefont
  {Singh}, \citenamefont {Zenesini}, \citenamefont {Morsch},\ and\
  \citenamefont {Arimondo}}]{PhysRevLett.99.220403}%
  \BibitemOpen
  \bibfield  {author} {\bibinfo {author} {\bibfnamefont {H.}~\bibnamefont
  {Lignier}}, \bibinfo {author} {\bibfnamefont {C.}~\bibnamefont {Sias}},
  \bibinfo {author} {\bibfnamefont {D.}~\bibnamefont {Ciampini}}, \bibinfo
  {author} {\bibfnamefont {Y.}~\bibnamefont {Singh}}, \bibinfo {author}
  {\bibfnamefont {A.}~\bibnamefont {Zenesini}}, \bibinfo {author}
  {\bibfnamefont {O.}~\bibnamefont {Morsch}},\ and\ \bibinfo {author}
  {\bibfnamefont {E.}~\bibnamefont {Arimondo}},\ }\bibfield  {title} {\bibinfo
  {title} {Dynamical control of matter-wave tunneling in periodic potentials},\
  }\href {https://doi.org/10.1103/PhysRevLett.99.220403} {\bibfield  {journal}
  {\bibinfo  {journal} {Phys. Rev. Lett.}\ }\textbf {\bibinfo {volume} {99}},\
  \bibinfo {pages} {220403} (\bibinfo {year} {2007})}\BibitemShut {NoStop}%
\bibitem [{\citenamefont {Lindner}\ \emph {et~al.}(2011)\citenamefont
  {Lindner}, \citenamefont {Refael},\ and\ \citenamefont
  {Galitski}}]{Lindner_2011}%
  \BibitemOpen
  \bibfield  {author} {\bibinfo {author} {\bibfnamefont {N.~H.}\ \bibnamefont
  {Lindner}}, \bibinfo {author} {\bibfnamefont {G.}~\bibnamefont {Refael}},\
  and\ \bibinfo {author} {\bibfnamefont {V.}~\bibnamefont {Galitski}},\
  }\bibfield  {title} {\bibinfo {title} {Floquet topological insulator in
  semiconductor quantum wells},\ }\href {https://doi.org/10.1038/nphys1926}
  {\bibfield  {journal} {\bibinfo  {journal} {Nature Physics}\ }\textbf
  {\bibinfo {volume} {7}},\ \bibinfo {pages} {490–495} (\bibinfo {year}
  {2011})}\BibitemShut {NoStop}%
\bibitem [{\citenamefont {Jotzu}\ \emph {et~al.}(2014)\citenamefont {Jotzu},
  \citenamefont {Messer}, \citenamefont {Desbuquois}, \citenamefont {Lebrat},
  \citenamefont {Uehlinger}, \citenamefont {Greif},\ and\ \citenamefont
  {Esslinger}}]{Jotzu_2014}%
  \BibitemOpen
  \bibfield  {author} {\bibinfo {author} {\bibfnamefont {G.}~\bibnamefont
  {Jotzu}}, \bibinfo {author} {\bibfnamefont {M.}~\bibnamefont {Messer}},
  \bibinfo {author} {\bibfnamefont {R.}~\bibnamefont {Desbuquois}}, \bibinfo
  {author} {\bibfnamefont {M.}~\bibnamefont {Lebrat}}, \bibinfo {author}
  {\bibfnamefont {T.}~\bibnamefont {Uehlinger}}, \bibinfo {author}
  {\bibfnamefont {D.}~\bibnamefont {Greif}},\ and\ \bibinfo {author}
  {\bibfnamefont {T.}~\bibnamefont {Esslinger}},\ }\bibfield  {title} {\bibinfo
  {title} {Experimental realization of the topological haldane model with
  ultracold fermions},\ }\href {https://doi.org/10.1038/nature13915} {\bibfield
   {journal} {\bibinfo  {journal} {Nature}\ }\textbf {\bibinfo {volume}
  {515}},\ \bibinfo {pages} {237–240} (\bibinfo {year} {2014})}\BibitemShut
  {NoStop}%
\bibitem [{\citenamefont {Mukherjee}\ and\ \citenamefont
  {Dutta}(2009)}]{Mukherjee_2009}%
  \BibitemOpen
  \bibfield  {author} {\bibinfo {author} {\bibfnamefont {V.}~\bibnamefont
  {Mukherjee}}\ and\ \bibinfo {author} {\bibfnamefont {A.}~\bibnamefont
  {Dutta}},\ }\bibfield  {title} {\bibinfo {title} {Effects of interference in
  the dynamics of a spin- 1/2 transversexychain driven periodically through
  quantum critical points},\ }\href
  {https://doi.org/10.1088/1742-5468/2009/05/p05005} {\bibfield  {journal}
  {\bibinfo  {journal} {Journal of Statistical Mechanics: Theory and
  Experiment}\ }\textbf {\bibinfo {volume} {2009}},\ \bibinfo {pages} {P05005}
  (\bibinfo {year} {2009})}\BibitemShut {NoStop}%
\bibitem [{\citenamefont {Das}(2010)}]{Das_2010}%
  \BibitemOpen
  \bibfield  {author} {\bibinfo {author} {\bibfnamefont {A.}~\bibnamefont
  {Das}},\ }\bibfield  {title} {\bibinfo {title} {Exotic freezing of response
  in a quantum many-body system},\ }\bibfield  {journal} {\bibinfo  {journal}
  {Physical Review B}\ }\textbf {\bibinfo {volume} {82}},\ \href
  {https://doi.org/10.1103/physrevb.82.172402} {10.1103/physrevb.82.172402}
  (\bibinfo {year} {2010})\BibitemShut {NoStop}%
\bibitem [{\citenamefont {Russomanno}\ \emph {et~al.}(2012)\citenamefont
  {Russomanno}, \citenamefont {Silva},\ and\ \citenamefont
  {Santoro}}]{Russomanno_2012}%
  \BibitemOpen
  \bibfield  {author} {\bibinfo {author} {\bibfnamefont {A.}~\bibnamefont
  {Russomanno}}, \bibinfo {author} {\bibfnamefont {A.}~\bibnamefont {Silva}},\
  and\ \bibinfo {author} {\bibfnamefont {G.~E.}\ \bibnamefont {Santoro}},\
  }\bibfield  {title} {\bibinfo {title} {Periodic steady regime and
  interference in a periodically driven quantum system},\ }\bibfield  {journal}
  {\bibinfo  {journal} {Physical Review Letters}\ }\textbf {\bibinfo {volume}
  {109}},\ \href {https://doi.org/10.1103/physrevlett.109.257201}
  {10.1103/physrevlett.109.257201} (\bibinfo {year} {2012})\BibitemShut
  {NoStop}%
\bibitem [{\citenamefont {Bastidas}\ \emph {et~al.}(2012)\citenamefont
  {Bastidas}, \citenamefont {Emary}, \citenamefont {Schaller},\ and\
  \citenamefont {Brandes}}]{Bastidas_2012}%
  \BibitemOpen
  \bibfield  {author} {\bibinfo {author} {\bibfnamefont {V.~M.}\ \bibnamefont
  {Bastidas}}, \bibinfo {author} {\bibfnamefont {C.}~\bibnamefont {Emary}},
  \bibinfo {author} {\bibfnamefont {G.}~\bibnamefont {Schaller}},\ and\
  \bibinfo {author} {\bibfnamefont {T.}~\bibnamefont {Brandes}},\ }\bibfield
  {title} {\bibinfo {title} {Nonequilibrium quantum phase transitions in the
  ising model},\ }\bibfield  {journal} {\bibinfo  {journal} {Physical Review
  A}\ }\textbf {\bibinfo {volume} {86}},\ \href
  {https://doi.org/10.1103/physreva.86.063627} {10.1103/physreva.86.063627}
  (\bibinfo {year} {2012})\BibitemShut {NoStop}%
\bibitem [{\citenamefont {Dunlap}\ and\ \citenamefont
  {Kenkre}(1986)}]{PhysRevB.34.3625}%
  \BibitemOpen
  \bibfield  {author} {\bibinfo {author} {\bibfnamefont {D.~H.}\ \bibnamefont
  {Dunlap}}\ and\ \bibinfo {author} {\bibfnamefont {V.~M.}\ \bibnamefont
  {Kenkre}},\ }\bibfield  {title} {\bibinfo {title} {Dynamic localization of a
  charged particle moving under the influence of an electric field},\ }\href
  {https://doi.org/10.1103/PhysRevB.34.3625} {\bibfield  {journal} {\bibinfo
  {journal} {Phys. Rev. B}\ }\textbf {\bibinfo {volume} {34}},\ \bibinfo
  {pages} {3625} (\bibinfo {year} {1986})}\BibitemShut {NoStop}%
\bibitem [{\citenamefont {Grossmann}\ \emph {et~al.}(1991)\citenamefont
  {Grossmann}, \citenamefont {Dittrich}, \citenamefont {Jung},\ and\
  \citenamefont {H\"anggi}}]{PhysRevLett.67.516}%
  \BibitemOpen
  \bibfield  {author} {\bibinfo {author} {\bibfnamefont {F.}~\bibnamefont
  {Grossmann}}, \bibinfo {author} {\bibfnamefont {T.}~\bibnamefont {Dittrich}},
  \bibinfo {author} {\bibfnamefont {P.}~\bibnamefont {Jung}},\ and\ \bibinfo
  {author} {\bibfnamefont {P.}~\bibnamefont {H\"anggi}},\ }\bibfield  {title}
  {\bibinfo {title} {Coherent destruction of tunneling},\ }\href
  {https://doi.org/10.1103/PhysRevLett.67.516} {\bibfield  {journal} {\bibinfo
  {journal} {Phys. Rev. Lett.}\ }\textbf {\bibinfo {volume} {67}},\ \bibinfo
  {pages} {516} (\bibinfo {year} {1991})}\BibitemShut {NoStop}%
\bibitem [{\citenamefont {{Kierig}}\ \emph {et~al.}(2008)\citenamefont
  {{Kierig}}, \citenamefont {{Schnorrberger}}, \citenamefont {{Schietinger}},
  \citenamefont {{Tomkovic}},\ and\ \citenamefont
  {{Oberthaler}}}]{2008PhRvL.100s0405K}%
  \BibitemOpen
  \bibfield  {author} {\bibinfo {author} {\bibfnamefont {E.}~\bibnamefont
  {{Kierig}}}, \bibinfo {author} {\bibfnamefont {U.}~\bibnamefont
  {{Schnorrberger}}}, \bibinfo {author} {\bibfnamefont {A.}~\bibnamefont
  {{Schietinger}}}, \bibinfo {author} {\bibfnamefont {J.}~\bibnamefont
  {{Tomkovic}}},\ and\ \bibinfo {author} {\bibfnamefont {M.~K.}\ \bibnamefont
  {{Oberthaler}}},\ }\bibfield  {title} {\bibinfo {title} {{Single-Particle
  Tunneling in Strongly Driven Double-Well Potentials}},\ }\href
  {https://doi.org/10.1103/PhysRevLett.100.190405} {\bibfield  {journal}
  {\bibinfo  {journal} {\prl}\ }\textbf {\bibinfo {volume} {100}},\ \bibinfo
  {eid} {190405} (\bibinfo {year} {2008})},\ \Eprint
  {https://arxiv.org/abs/0803.1406} {arXiv:0803.1406 [quant-ph]} \BibitemShut
  {NoStop}%
\bibitem [{\citenamefont {{Eckardt}}\ \emph {et~al.}(2005)\citenamefont
  {{Eckardt}}, \citenamefont {{Weiss}},\ and\ \citenamefont
  {{Holthaus}}}]{2005PhRvL..95z0404E}%
  \BibitemOpen
  \bibfield  {author} {\bibinfo {author} {\bibfnamefont {A.}~\bibnamefont
  {{Eckardt}}}, \bibinfo {author} {\bibfnamefont {C.}~\bibnamefont {{Weiss}}},\
  and\ \bibinfo {author} {\bibfnamefont {M.}~\bibnamefont {{Holthaus}}},\
  }\bibfield  {title} {\bibinfo {title} {{Superfluid-Insulator Transition in a
  Periodically Driven Optical Lattice}},\ }\href
  {https://doi.org/10.1103/PhysRevLett.95.260404} {\bibfield  {journal}
  {\bibinfo  {journal} {\prl}\ }\textbf {\bibinfo {volume} {95}},\ \bibinfo
  {eid} {260404} (\bibinfo {year} {2005})},\ \Eprint
  {https://arxiv.org/abs/cond-mat/0601020} {arXiv:cond-mat/0601020
  [cond-mat.other]} \BibitemShut {NoStop}%
\bibitem [{\citenamefont {{Sillanp{\"a}{\"a}}}\ \emph
  {et~al.}(2006)\citenamefont {{Sillanp{\"a}{\"a}}}, \citenamefont
  {{Lehtinen}}, \citenamefont {{Paila}}, \citenamefont {{Makhlin}},\ and\
  \citenamefont {{Hakonen}}}]{2006PhRvL..96r7002S}%
  \BibitemOpen
  \bibfield  {author} {\bibinfo {author} {\bibfnamefont {M.}~\bibnamefont
  {{Sillanp{\"a}{\"a}}}}, \bibinfo {author} {\bibfnamefont {T.}~\bibnamefont
  {{Lehtinen}}}, \bibinfo {author} {\bibfnamefont {A.}~\bibnamefont {{Paila}}},
  \bibinfo {author} {\bibfnamefont {Y.}~\bibnamefont {{Makhlin}}},\ and\
  \bibinfo {author} {\bibfnamefont {P.}~\bibnamefont {{Hakonen}}},\ }\bibfield
  {title} {\bibinfo {title} {{Continuous-Time Monitoring of Landau-Zener
  Interference in a Cooper-Pair Box}},\ }\href
  {https://doi.org/10.1103/PhysRevLett.96.187002} {\bibfield  {journal}
  {\bibinfo  {journal} {\prl}\ }\textbf {\bibinfo {volume} {96}},\ \bibinfo
  {eid} {187002} (\bibinfo {year} {2006})},\ \Eprint
  {https://arxiv.org/abs/cond-mat/0510559} {arXiv:cond-mat/0510559
  [cond-mat.mes-hall]} \BibitemShut {NoStop}%
\bibitem [{\citenamefont {Lieb}\ \emph {et~al.}(1961)\citenamefont {Lieb},
  \citenamefont {Schultz},\ and\ \citenamefont {Mattis}}]{LIEB1961407}%
  \BibitemOpen
  \bibfield  {author} {\bibinfo {author} {\bibfnamefont {E.}~\bibnamefont
  {Lieb}}, \bibinfo {author} {\bibfnamefont {T.}~\bibnamefont {Schultz}},\ and\
  \bibinfo {author} {\bibfnamefont {D.}~\bibnamefont {Mattis}},\ }\bibfield
  {title} {\bibinfo {title} {Two soluble models of an antiferromagnetic
  chain},\ }\href
  {https://doi.org/https://doi.org/10.1016/0003-4916(61)90115-4} {\bibfield
  {journal} {\bibinfo  {journal} {Annals of Physics}\ }\textbf {\bibinfo
  {volume} {16}},\ \bibinfo {pages} {407 } (\bibinfo {year}
  {1961})}\BibitemShut {NoStop}%
\bibitem [{\citenamefont {Dziarmaga}(2005{\natexlab{b}})}]{Dziarmaga_2005}%
  \BibitemOpen
  \bibfield  {author} {\bibinfo {author} {\bibfnamefont {J.}~\bibnamefont
  {Dziarmaga}},\ }\bibfield  {title} {\bibinfo {title} {Dynamics of a quantum
  phase transition: Exact solution of the quantum ising model},\ }\bibfield
  {journal} {\bibinfo  {journal} {Physical Review Letters}\ }\textbf {\bibinfo
  {volume} {95}},\ \href {https://doi.org/10.1103/physrevlett.95.245701}
  {10.1103/physrevlett.95.245701} (\bibinfo {year}
  {2005}{\natexlab{b}})\BibitemShut {NoStop}%
\bibitem [{\citenamefont {Bhattacharyya}\ \emph {et~al.}(2012)\citenamefont
  {Bhattacharyya}, \citenamefont {Das},\ and\ \citenamefont
  {Dasgupta}}]{Bhattacharyya_2012}%
  \BibitemOpen
  \bibfield  {author} {\bibinfo {author} {\bibfnamefont {S.}~\bibnamefont
  {Bhattacharyya}}, \bibinfo {author} {\bibfnamefont {A.}~\bibnamefont {Das}},\
  and\ \bibinfo {author} {\bibfnamefont {S.}~\bibnamefont {Dasgupta}},\
  }\bibfield  {title} {\bibinfo {title} {Transverse ising chain under periodic
  instantaneous quenches: Dynamical many-body freezing and emergence of slow
  solitary oscillations},\ }\bibfield  {journal} {\bibinfo  {journal} {Physical
  Review B}\ }\textbf {\bibinfo {volume} {86}},\ \href
  {https://doi.org/10.1103/physrevb.86.054410} {10.1103/physrevb.86.054410}
  (\bibinfo {year} {2012})\BibitemShut {NoStop}%
\bibitem [{\citenamefont {Ashhab}\ \emph {et~al.}(2007)\citenamefont {Ashhab},
  \citenamefont {Johansson}, \citenamefont {Zagoskin},\ and\ \citenamefont
  {Nori}}]{Ashhab_2007}%
  \BibitemOpen
  \bibfield  {author} {\bibinfo {author} {\bibfnamefont {S.}~\bibnamefont
  {Ashhab}}, \bibinfo {author} {\bibfnamefont {J.~R.}\ \bibnamefont
  {Johansson}}, \bibinfo {author} {\bibfnamefont {A.~M.}\ \bibnamefont
  {Zagoskin}},\ and\ \bibinfo {author} {\bibfnamefont {F.}~\bibnamefont
  {Nori}},\ }\bibfield  {title} {\bibinfo {title} {Two-level systems driven by
  large-amplitude fields},\ }\bibfield  {journal} {\bibinfo  {journal}
  {Physical Review A}\ }\textbf {\bibinfo {volume} {75}},\ \href
  {https://doi.org/10.1103/physreva.75.063414} {10.1103/physreva.75.063414}
  (\bibinfo {year} {2007})\BibitemShut {NoStop}%
\bibitem [{\citenamefont {Barouch}\ \emph {et~al.}(1970)\citenamefont
  {Barouch}, \citenamefont {McCoy},\ and\ \citenamefont
  {Dresden}}]{PhysRevA.2.1075}%
  \BibitemOpen
  \bibfield  {author} {\bibinfo {author} {\bibfnamefont {E.}~\bibnamefont
  {Barouch}}, \bibinfo {author} {\bibfnamefont {B.~M.}\ \bibnamefont {McCoy}},\
  and\ \bibinfo {author} {\bibfnamefont {M.}~\bibnamefont {Dresden}},\
  }\bibfield  {title} {\bibinfo {title} {Statistical mechanics of the
  $\mathrm{XY}$ model. i},\ }\href {https://doi.org/10.1103/PhysRevA.2.1075}
  {\bibfield  {journal} {\bibinfo  {journal} {Phys. Rev. A}\ }\textbf {\bibinfo
  {volume} {2}},\ \bibinfo {pages} {1075} (\bibinfo {year} {1970})}\BibitemShut
  {NoStop}%
\bibitem [{\citenamefont {Heyl}(2018)}]{Heyl_2018}%
  \BibitemOpen
  \bibfield  {author} {\bibinfo {author} {\bibfnamefont {M.}~\bibnamefont
  {Heyl}},\ }\bibfield  {title} {\bibinfo {title} {Dynamical quantum phase
  transitions: a review},\ }\href {https://doi.org/10.1088/1361-6633/aaaf9a}
  {\bibfield  {journal} {\bibinfo  {journal} {Reports on Progress in Physics}\
  }\textbf {\bibinfo {volume} {81}},\ \bibinfo {pages} {054001} (\bibinfo
  {year} {2018})}\BibitemShut {NoStop}%
\bibitem [{\citenamefont {Yang}\ \emph {et~al.}(2019)\citenamefont {Yang},
  \citenamefont {Zhou}, \citenamefont {Ma}, \citenamefont {Kong}, \citenamefont
  {Wang}, \citenamefont {Qin}, \citenamefont {Rong}, \citenamefont {Wang},
  \citenamefont {Shi}, \citenamefont {Gong},\ and\ \citenamefont
  {et~al.}}]{Yang_2019}%
  \BibitemOpen
  \bibfield  {author} {\bibinfo {author} {\bibfnamefont {K.}~\bibnamefont
  {Yang}}, \bibinfo {author} {\bibfnamefont {L.}~\bibnamefont {Zhou}}, \bibinfo
  {author} {\bibfnamefont {W.}~\bibnamefont {Ma}}, \bibinfo {author}
  {\bibfnamefont {X.}~\bibnamefont {Kong}}, \bibinfo {author} {\bibfnamefont
  {P.}~\bibnamefont {Wang}}, \bibinfo {author} {\bibfnamefont {X.}~\bibnamefont
  {Qin}}, \bibinfo {author} {\bibfnamefont {X.}~\bibnamefont {Rong}}, \bibinfo
  {author} {\bibfnamefont {Y.}~\bibnamefont {Wang}}, \bibinfo {author}
  {\bibfnamefont {F.}~\bibnamefont {Shi}}, \bibinfo {author} {\bibfnamefont
  {J.}~\bibnamefont {Gong}},\ and\ \bibinfo {author} {\bibnamefont {et~al.}},\
  }\bibfield  {title} {\bibinfo {title} {Floquet dynamical quantum phase
  transitions},\ }\bibfield  {journal} {\bibinfo  {journal} {Physical Review
  B}\ }\textbf {\bibinfo {volume} {100}},\ \href
  {https://doi.org/10.1103/physrevb.100.085308} {10.1103/physrevb.100.085308}
  (\bibinfo {year} {2019})\BibitemShut {NoStop}%
\bibitem [{\citenamefont {Gritsev}\ and\ \citenamefont
  {Polkovnikov}(2017)}]{Gritsev_2017}%
  \BibitemOpen
  \bibfield  {author} {\bibinfo {author} {\bibfnamefont {V.}~\bibnamefont
  {Gritsev}}\ and\ \bibinfo {author} {\bibfnamefont {A.}~\bibnamefont
  {Polkovnikov}},\ }\bibfield  {title} {\bibinfo {title} {Integrable floquet
  dynamics},\ }\bibfield  {journal} {\bibinfo  {journal} {SciPost Physics}\
  }\textbf {\bibinfo {volume} {2}},\ \href
  {https://doi.org/10.21468/scipostphys.2.3.021} {10.21468/scipostphys.2.3.021}
  (\bibinfo {year} {2017})\BibitemShut {NoStop}%
\bibitem [{\citenamefont {Zhang}\ \emph {et~al.}(2016)\citenamefont {Zhang},
  \citenamefont {Khemani},\ and\ \citenamefont {Huse}}]{Zhang_2016}%
  \BibitemOpen
  \bibfield  {author} {\bibinfo {author} {\bibfnamefont {L.}~\bibnamefont
  {Zhang}}, \bibinfo {author} {\bibfnamefont {V.}~\bibnamefont {Khemani}},\
  and\ \bibinfo {author} {\bibfnamefont {D.~A.}\ \bibnamefont {Huse}},\
  }\bibfield  {title} {\bibinfo {title} {A floquet model for the many-body
  localization transition},\ }\bibfield  {journal} {\bibinfo  {journal}
  {Physical Review B}\ }\textbf {\bibinfo {volume} {94}},\ \href
  {https://doi.org/10.1103/physrevb.94.224202} {10.1103/physrevb.94.224202}
  (\bibinfo {year} {2016})\BibitemShut {NoStop}%
\end{thebibliography}%

\end{document}